\documentclass[a4paper, amsfonts, amssymb, amsmath, preprint, showkeys, nofootinbib, twoside, superscriptaddress]{revtex4-2}
\usepackage[english]{babel}
\usepackage[utf8]{inputenc}
\usepackage[colorinlistoftodos, color=green!40, prependcaption]{todonotes}
\usepackage{amsthm}
\usepackage{mathtools}
\usepackage{physics}
\usepackage{xcolor}
\usepackage{graphicx}
\usepackage[left=23mm,right=13mm,top=35mm,columnsep=15pt]{geometry} 
\usepackage{adjustbox}
\usepackage{placeins}
\usepackage[T1]{fontenc}
\usepackage{lipsum}
\usepackage{csquotes}
\usepackage[pdftex, pdftitle={Article}, pdfauthor={Author}]{hyperref} 
\usepackage{caption}
\usepackage{subcaption}
\usepackage{xcolor}
\usepackage{tablefootnote}
\usepackage{amsmath}
\begin{document}
\title{The diffusion-driven orthorhombic to tetragonal transition in YBa$_2$Cu$_3$O$_7$ derived with a machine learning interatomic potential}

\author{Davide Gambino}
    \email[Correspondence email address: ]{davide.gambino@liu.se}
    \affiliation{Theoretical Physics Division, Department of Physics, Chemistry, and Biology (IFM), Linköping University, Linköping 58183, Sweden}
    \affiliation{Department of Physics, P.O. Box 43, FI-00014 University of Helsinki, Finland}
\author{Niccolò Di Eugenio}
    \affiliation{Department of Applied Science and Technology, Politecnico di Torino, I-10129 Torino, Italy}
    \affiliation{Istituto Nazionale di Fisica Nucleare, Sezione di Torino, I-10125 Torino, Italy}
\author{Jesper Byggm\"astar}
     \affiliation{Department of Physics, P.O. Box 43, FI-00014 University of Helsinki, Finland}
\author{Johan Klarbring}
    \affiliation{Theoretical Physics Division, Department of Physics, Chemistry, and Biology (IFM), Linköping University, Linköping 58183, Sweden}
\author{Daniele Torsello}
    \affiliation{Department of Applied Science and Technology, Politecnico di Torino, I-10129 Torino, Italy}
    \affiliation{Istituto Nazionale di Fisica Nucleare, Sezione di Torino, I-10125 Torino, Italy}
\author{Flyura Djurabekova}
     \affiliation{Department of Physics, P.O. Box 43, FI-00014 University of Helsinki, Finland}
     \affiliation{Helsinki Institute of Physics, Helsinki, Finland}
\author{Francesco Laviano}
    \affiliation{Department of Applied Science and Technology, Politecnico di Torino, I-10129 Torino, Italy}
    \affiliation{Istituto Nazionale di Fisica Nucleare, Sezione di Torino, I-10125 Torino, Italy}
\date{\today} 

\begin{abstract}
Defects in high temperature superconductors such as YBa$_2$Cu$_3$O$_7$ (YBCO) critically influence their superconducting behavior, as they substantially degrade or even suppress superconductivity. 
With the renewed interest in cuprates for next-generation superconducting magnets operating in radiation-harsh environments such as fusion reactors and particle accelerators, accurate atomistic modeling of defects and their dynamics has become essential. 
Here, we present a general-purpose machine-learning interatomic potential for YBCO, based on the Atomic Cluster Expansion (ACE) method and trained on Density Functional Theory (DFT) data, with particular emphasis on defects and their diffusion mechanisms. 
The potential is validated against DFT calculations of ground-state properties, defect formation energies of oxygen Frenkel pairs and diffusion barriers for their formation. 
Remarkably, the potential captures the diffusion-driven orthorhombic to tetragonal transition at elevated temperatures, a transformation that is difficult to describe with empirical potentials, elucidating how the formation of oxygen Frenkel pairs in the basal plane governs this order-disorder transition.
The ACE potential introduced here enables large-scale, predictive atomistic simulations of defect dynamics and transport processes in YBCO, providing a powerful tool to explore its stability, performance, and functionality under realistic operating conditions.
Moreover, this work proves that machine learning interatomic potentials are suitable for studies of quaternary oxides with complex chemistry.
\end{abstract}


\maketitle

\section{Introduction} \label{sec:introduction}
Rare-earth barium cuprates (REBCOs) are subject to renewed interest \cite{HTS_applications} following recent technological developments in the fabrication of high-temperature superconducting (HTS) tapes \cite{2gHTS_2014,2gHTS_2021} 
and their proposed employment in compact \cite{ARC_original,SPARC_magnet,TE,Commercial_fusion,HTS_tokamak_2024} and standard fusion reactors \cite{DEMO_CS,DEMO_HTS}, as well as particle accelerators \cite{Muon_colliders,HTS_PSI,HTS_CERN}.
Since their discovery \cite{HTS_discovery}, REBCOs have proven to be extremely challenging materials to investigate both experimentally and theoretically.
The coupling mechanism driving superconductivity in these materials remains debated \cite{Unconventional_superconductivity_I,Unconventional_superconductivity_II,Unconventional_superconductivity_III,Unconventional_superconductivity_IV,Unconventional_superconductivity_V}, and all their properties are intimately linked with the stoichiometry of the oxygen sublattice. 
In YBa$_2$Cu$_3$O$_{7-x}$ (YBCO), for example, varying $x$ drives the system from an antiferromagnetic insulator ($x\approx 1$) to a strange metal/superconductor ($x\approx 0$) \cite{YBCO_phase_diagram_new}. 

In addition, the synthesis of REBCOs is plagued by defects: single crystals of these materials are difficult to produce and are characterized by the presence of extended defects such as twin boundaries \cite{YBCO_twinning_I} and a certain degree of disorder in the oxygen sublattice \cite{YBCO_O_disorder_RBS}, with beneficial (vortex pinning) and detrimental (Cooper pair scattering) effects depending on the type of defect \cite{Defect_scattering_in_superconductors,HTS_pinning_centers_review,obradors_pin_2024,puig_impact_2024,ruiz_critical_2026}.
Oxygen vacancies in the CuO chains in the basal plane are known to introduce holes in the CuO$_2$ planes \cite{YBCO_hole_doping_O_deficiency}, and their amount governs the superconducting critical properties \cite{YBCO_phase_diagram}.
Indeed, it is observed that annealing of irradiated samples partly recovers the pristine critical superconducting properties even at cryogenic temperatures \cite{Legris_annealing_1993,Unterrainer_annealing,GdBCO,Fischer_annealing_2025}, strongly indicating that the O defect diffusion and recombination is an extremely important effect in these materials \cite{Mundet_2020}. 
Detailed understanding of the defect formation process is of utmost importance for technological applications in radiation harsh environments, and impact of different types of defects on the superconducting properties should be understood to improve HTS device design \cite{Torsello_ARC_magnets}.
Because of experimental limitations in reproducing operational conditions of, e.g., fusion reactors \cite{Torsello_YBCO_radiation_damage}, computational modeling with first principles and atomistic methods can guide the development of devices by giving insights on the microscopic mechanisms of degradation and on possible recovery mechanisms \cite{HTS_fusion_roadmap_2025}.

Disorder in the oxygen sublattice is also responsible for the orthorhombic to tetragonal transition occurring at high temperatures over a broad range of stoichiometries \cite{expTransition,ExpTransitionAndSC,expTransition_vs_O_content}.
Neutron diffraction studies of YBCO \cite{expTransition,ExpTransitionAndSC} have shown that the mechanism governing this transition consists of a rearrangement of the occupation of the O sites in the basal plane, with O atoms from the CuO chains moving to the available sites in between the chains.
This disorder in the basal plane leads to a lack of preferential direction in the $a$ and $b$ directions, driving the transition to a higher-symmetry crystal structure.

The orthorhombic-to-tetragonal transition has also been observed in molecular dynamics (MD) simulations by Chaplot \cite{Chaplot_potential_transition}, using a simple interatomic potential specifically tuned to reproduce the experimental transition temperature \cite{Chaplot_potential_phonons,Chaplot_potential_transition,Chaplot_potential_transition_and_phonons}. 
A more general classical interatomic potential was later developed by Gray et al. \cite{GrayPotential} for studying radiation damage and defect properties \cite{GrayPotential,Torsello_YBCO_radiation_damage,Dickson_TDE}. This potential combines Buckingham and Coulomb terms, with parameters fitted to density functional theory (DFT) data; however, it exhibits some limitations, for instance, it fails to capture the high-temperature structural transition.

Machine Learning Potentials (MLPs) \cite{ML_general_I,ML_general_II,ML_general_III,ML_general_IV}  trained on DFT calculations offer a powerful route to overcome such limitations. 
However, specialized MLPs for systems with many elements and complex chemistry are quite rare \cite{Pacemaker_PRM,ACE_NaZrSiPO,ACE_NiCrFLiBe}, and even further challenges are introduced by the modeling of defects \cite{ML_W_defs,ACE_CuZr_defs,ML_discrepancies_defs,ML_Cu_Al_Ni_defs,ML_MPIE_defs}.
YBCO, with its large unit cell, many inequivalent atomic sites and its diffusion-driven, order-disorder structural transition, provides a challenging testbed for assessing the capabilities of MLPs to describe diverse bonding in the same compound and defect dynamics. 
In this work, we develop a MLP based on the Atomic Cluster Expansion (ACE) method \cite{ACE_Drautz_I,ACE_Drautz_II} trained on DFT data including defects of all species. 
The ACE potential improves upon the classical potential of Gray et el. \cite{GrayPotential} across a wide range of properties and, crucially, reproduces the orthorhombic-to-tetragonal transition at elevated temperatures. 
Our simulations reveal that this transition is entropy-driven, facilitated by the decreasing formation energy of oxygen Frenkel pairs (FPs) in the basal plane due to thermal expansion. 
Our results demonstrate the capability of MLPs to capture complex defect physics in quaternary oxide compounds, paving the way for predictive simulations of their behavior under radiation harsh conditions.

\section{Results} \label{sec:results}

\subsection{DFT reference data set and training of the ACE potential}

\begin{figure}
    \centering
    \includegraphics[width=0.85\columnwidth]{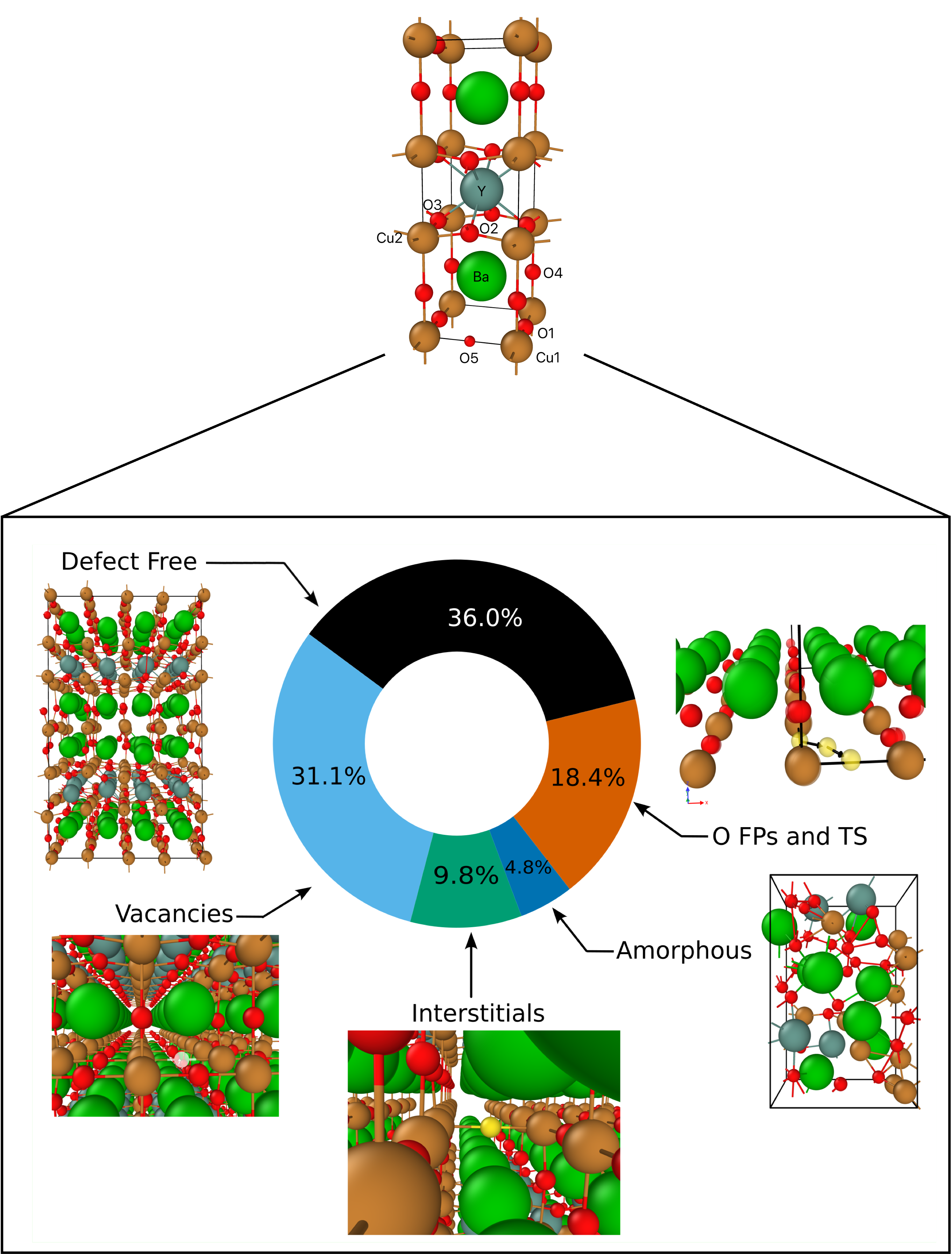}
    \caption{Unit cell of YBCO with its inequivalent crystallographic positions and the most stable interstitial position for O atoms (O5 position), together with a visual representation of the dataset with example structural models generated with OVITO \cite{ovito}.}
    \label{fig:1_combo}
\end{figure}

The training dataset was constructed to capture both equilibrium and non-equilibrium environments relevant to defect formation and diffusion in YBCO. 
Configurations were first generated from MD simulations (416 atoms) over a wide temperature range for the defect-free structure and for structures containing all possible vacancies and interstitials, using a preliminary ACE potential trained on low-accuracy ab initio MD and defect relaxations taken from Ref. \cite{GdBCO}. 
Additional defect-free structures over a broad range of volumes from DFT relaxations were included, spanning compressed to highly expanded lattices. 
To ensure representation of disordered states, we also incorporated manually generated amorphous-like configurations (52 atoms) that exhibit liquid-like local environments. 
To better target the orthorhombic-to-tetragonal transition, the dataset was enriched with perturbed structures containing oxygen FPs (vacancies at each inequivalent oxygen site combined with O interstitials at the O5 site; see Fig. \ref{fig:1_combo}), as well as approximate transition-state (TS) geometries for FP formation. 
The diversity of these configurations, summarized visually in Fig. \ref{fig:1_combo}, ensures balanced coverage of the relevant defect physics and structural transitions.
The data set is composed of 1368 configurations (545267 atoms) in total, see Supplementary Material for further details.

\begin{figure}
    \centering
    \begin{subfigure}{\columnwidth}
         \centering
         \includegraphics[width=\columnwidth]{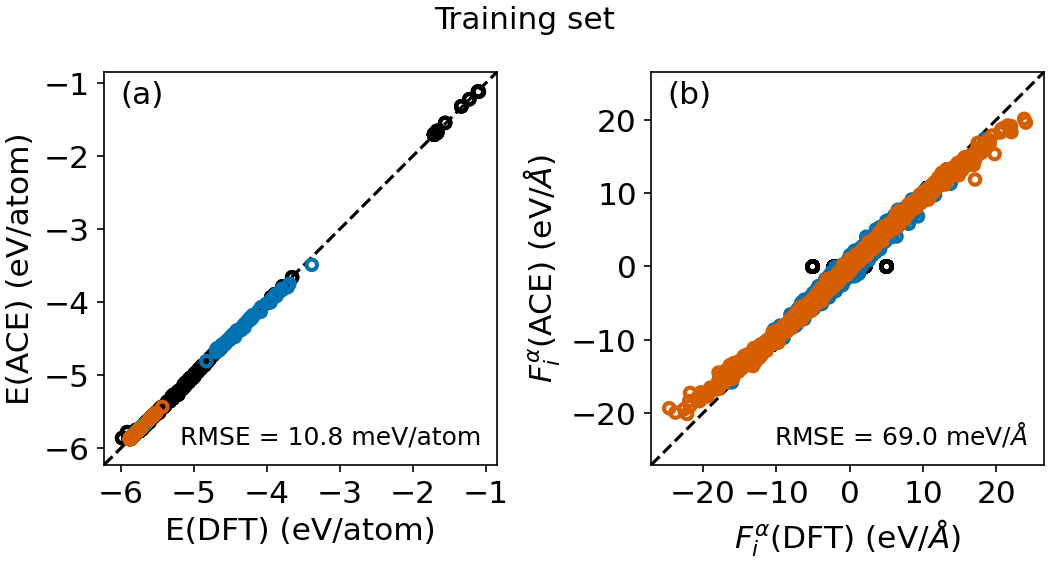}
    \end{subfigure}
    \begin{subfigure}{\columnwidth}
         \centering
         \includegraphics[width=\columnwidth]{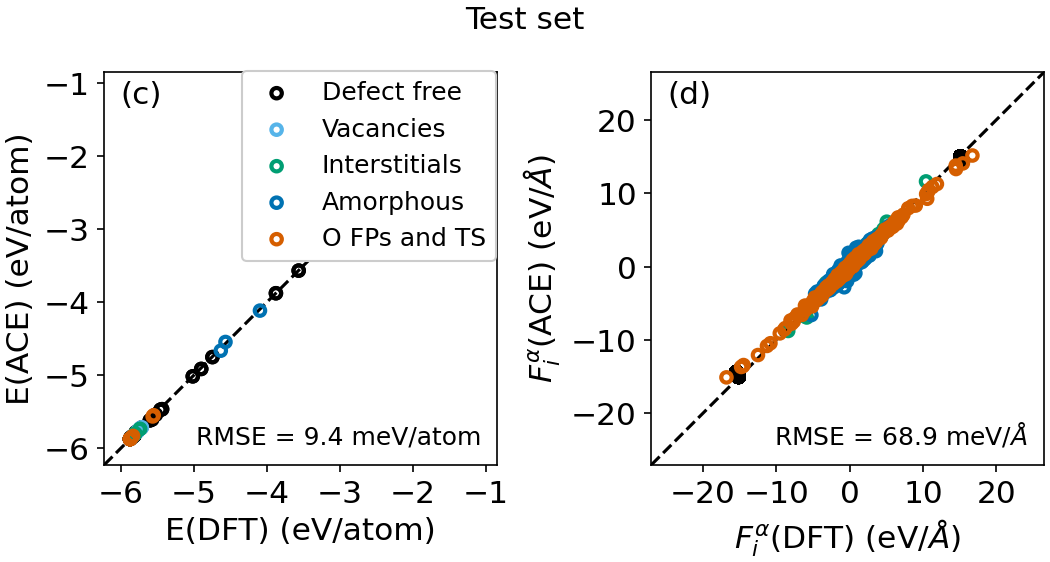}
    \end{subfigure}
    \caption{Parity plots of training (top row) and test (bottom row) sets of energies (left column) and forces (right column), with each subset represented with different colors (see legend for details). RMSEs are 10.8 meV/atom and 69.0 meV/\AA \; on energy and forces from the training set, and 9.4 meV/atom and 68.9 meV/\AA \; from the test set.} 
    \label{fig:3_pairplots}
\end{figure}

As shown in the parity plots (Fig. \ref{fig:3_pairplots}), the potential achieves high accuracy for both energies and forces, with RMSEs of 10 meV/atom and 70 meV/\AA, respectively, on both the training and test sets (the latter comprising 5\% of the full dataset, randomly selected).
A small set of off-diagonal points in Fig. \ref{fig:3_pairplots}b was traced to a DFT artifact arising during fixed-volume relaxation of the ideal structure.
After recalculating the problematic configuration, the ACE potential was found to predict energies and forces in much closer agreement with the corrected DFT values ($\approx 4$ meV/atom for energies and $\leq 15$ meV/\AA \; for forces). 
This demonstrates that the potential remains robust and accurate, even in the presence of such problematic configuration in the training set.

\subsection{Validation of the ACE potential}

\begin{figure}
    \centering
    \includegraphics[width=\columnwidth]{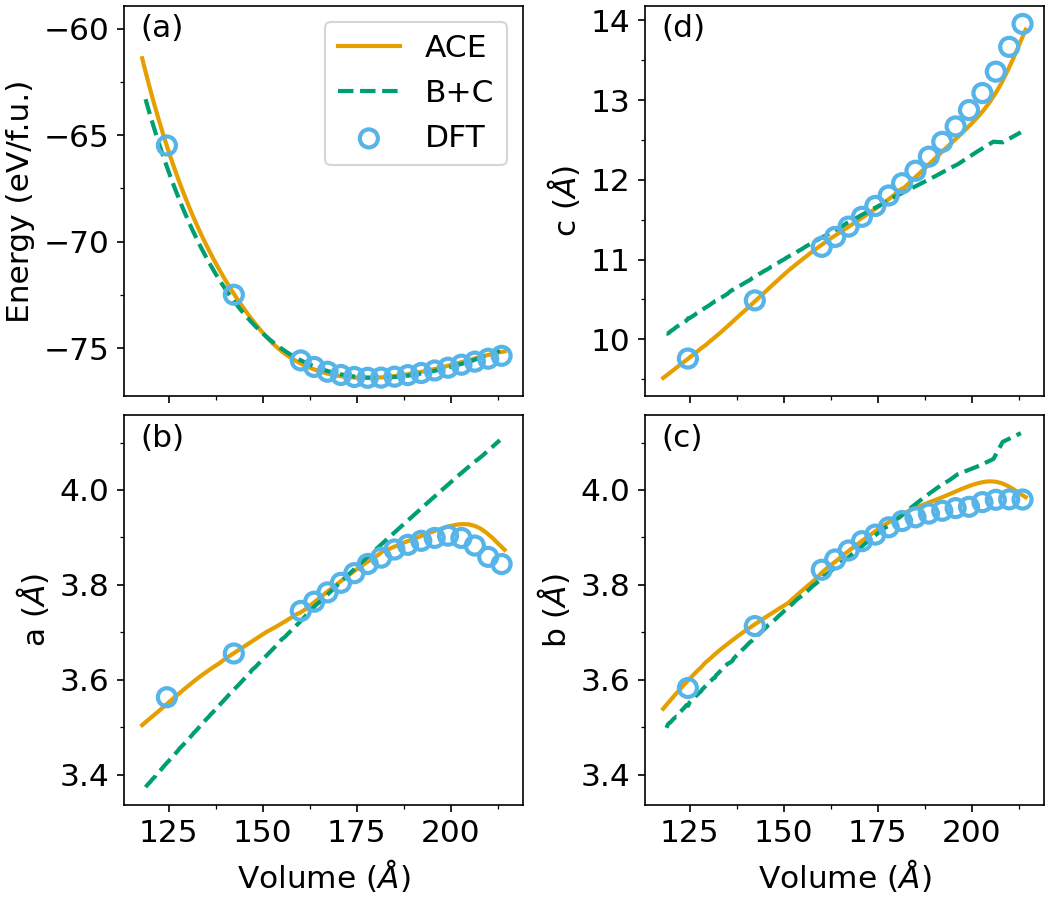}
    \caption{Equation of state (top left) and lattice parameters $a$, $b$, and $c$ (bottom left, bottom right, and top right, respectively) as a function of volume at 0 K from DFT (blue empty circles), the Buckingham+Coulomb potential (B+C, green dashed line), and the ACE potential (orange solid line). The B+C equation of state curve has been rigidly shifted vertically as to match the DFT minimum energy.}
    \label{fig:4_EOS}
\end{figure}

To validate the accuracy of the developed potential, we calculated the equation of state for YBCO using our ACE potential and compared it with results from  DFT and the Buckingham+Coulomb (B+C) classical potential of Gray et al. \cite{GrayPotential} (Fig. \ref{fig:4_EOS}a).
Both the ACE and the B+C potentials closely reproduce the DFT results, although the B+C curve has been shifted vertically here to match the minimum with the DFT results.
The superiority of the ACE potential becomes apparent when examining the $a$, $b$, and $c$ lattice parameters (Fig. \ref{fig:4_EOS} b-d): ACE accurately captures the slope of all lattice parameters around equilibrium and closely follows the non-monotonic trend at large volumes observed in DFT.
In contrast, the B+C potential overestimates the slope of the $a$ lattice parameter, underestimates that of the $c$ parameter, and fails to reproduce the large-volume behavior.
This failure of the B+C potential is probably due to the lack of flexibility of this model, inherited from the fixed charges assigned to each species.
ACE is able to capture local rearrangements occurring at volumes far from equilibrium, demonstrating its ability to reproduce the physical behavior of this challenging system.

\begin{figure}
    \centering
    \includegraphics[width=\columnwidth]{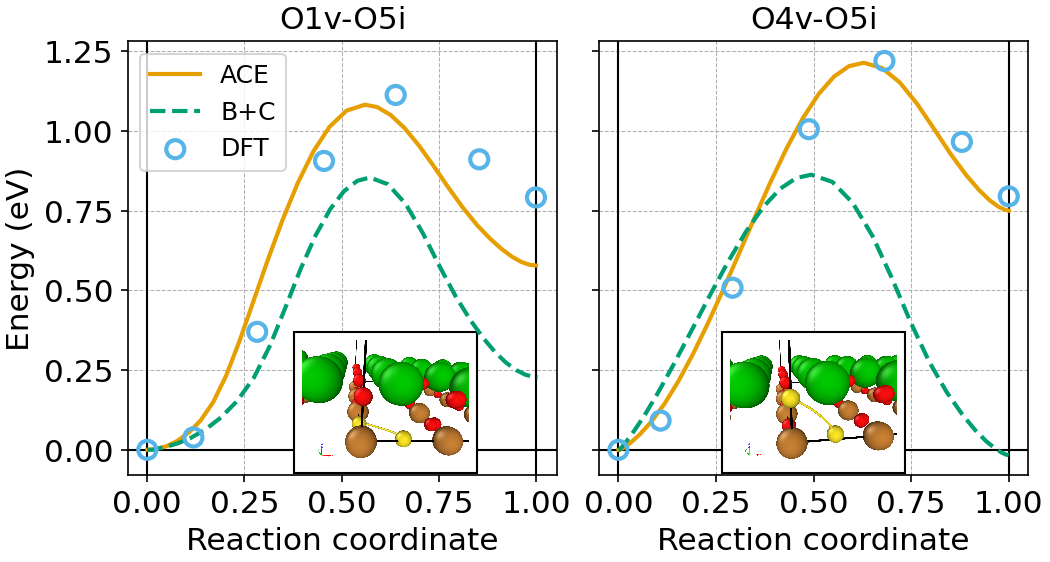}
    \caption{Energy barrier for the formation of the O FPs O1v-O5i (left) and O4v-O5i (right) from DFT (blue empty circles), ACE (orange solid lines), and the B+C potential (green dashed lines). The final energy corresponds to the formation energy of the FP. The figure insets in each panel show the migration path within the YBCO cell.}
    \label{fig:6_NEBs}
\end{figure}

We further validated the potential on defect energetics relevant to the orthorhombic-to-tetragonal transition, focusing on FPs formed by diffusion of O from the O1 and O4 sites to the O5 site (see Fig. \ref{fig:6_NEBs}).
The ACE potential reproduces the migration energies of both processes with high accuracy, within 20 meV of the DFT reference, substantially improving upon the B+C potential, which underestimates these barriers by 0.3-0.4 eV.
Formation energies are similarly well captured: for O4v–O5i, ACE predicts 0.77 eV versus 0.81 eV from DFT, while B+C incorrectly stabilizes this defect relative to the pristine structure. 
The O1v–O5i formation energy is slightly underestimated by ACE (0.62 vs 0.79 eV from DFT), but remains significantly closer than B+C ($\approx 0.5$ eV below DFT). 
This defect is intrinsically difficult to reproduce, as the local O coordination with its Cu first nearest neighbors is nearly identical to the defect-free environment. 
Nonetheless, errors of $\approx 0.2$ eV in defect formation energies are typical for MLPs \cite{ML_W_defs,ACE_CuZr_defs,ML_discrepancies_defs,ML_Cu_Al_Ni_defs,ML_MPIE_defs}, and the high accuracy in reproducing migration barriers underscores the reliability of the potential for the investigation of oxygen diffusion in YBCO.
Additional validation on lattice dynamics, mechanical properties, and O FPs energetics is shown in Supplementary Material.

\subsection{The orthorhombic to tetragonal transition}
\begin{figure}
    \centering
    \includegraphics[width=\columnwidth]{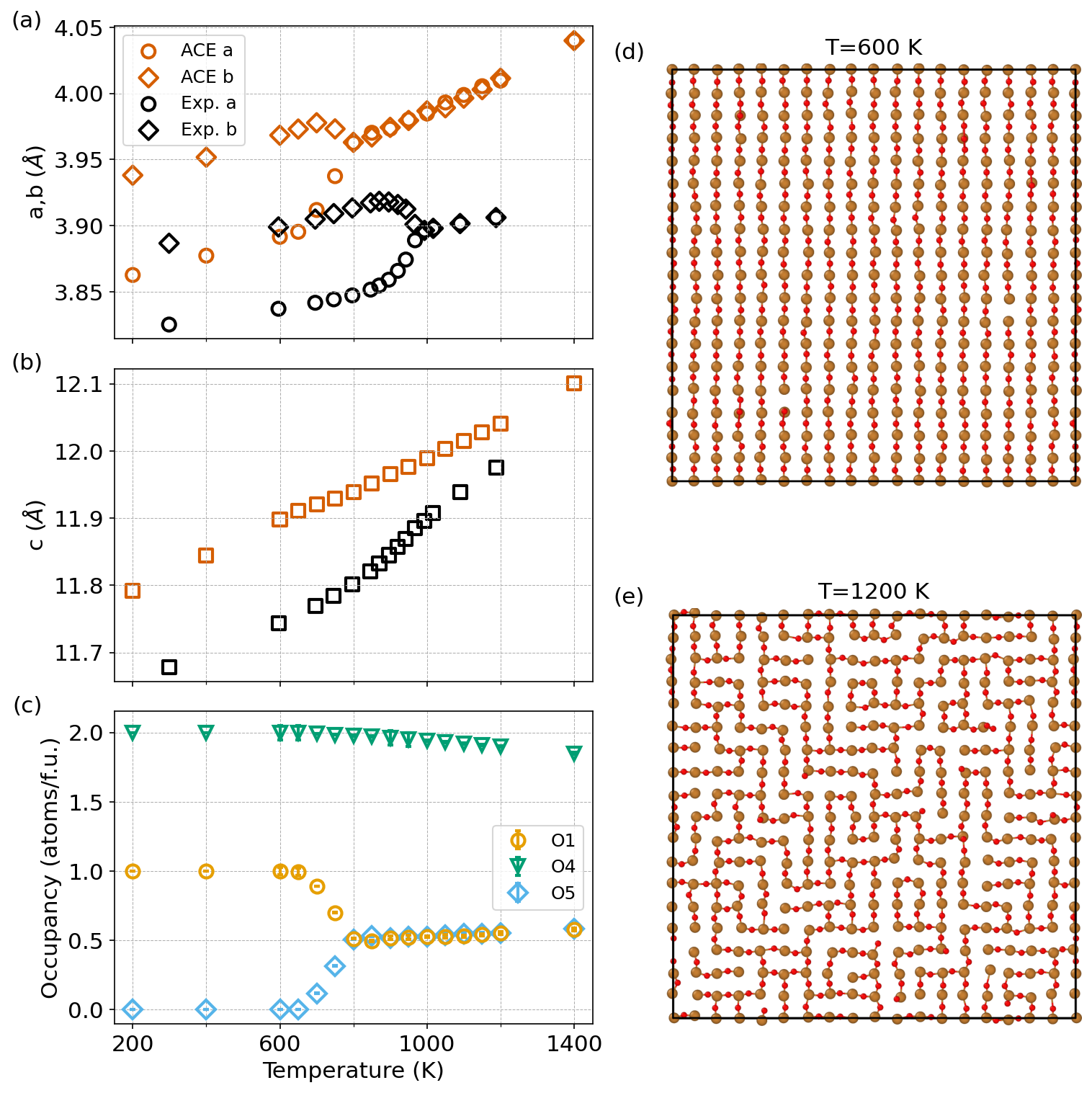}
    \caption{a) $a$ (circles) and $b$ (diamonds) lattice parameters as a function of temperature from ACE (orange) and experiments (black) \cite{expTransition}. b) $c$ lattice  parameter; c) Occupation of O1, O4, and O5 sites as a function of temperature. The figures on the right show snapshots of the basal plane d) below  and e) above the transition.}
    \label{fig:7_lps_vs_T}
\end{figure}

Capturing the orthorhombic to tetragonal transition in YBCO is particularly challenging for interatomic potentials, since it is driven by subtle diffusion processes and disorder in the oxygen sublattice.
Here, we show that the ACE potential is able to reproduce this transition by carrying out NPT simulations over a wide temperature range (200-1400 K) with large simulation cells ($\approx 25000$ atoms) up to 75 ns. 
The transition occurs spontaneously at approximately 800 K (See Fig. \ref{fig:7_lps_vs_T} a), with signs of a disordered orthorhombic phase already at 700 K, in reasonable agreement with the experimental transition observed at 1000 K \cite{expTransition}.
Although the simulations were run for unusually long times compared to typical MD studies, the structures at 700 and 750 K still exhibited incomplete convergence (see Supplementary Material for details on the lattice parameters).

The transition can be identified as an order-disorder transition, in which the linearity of the CuO chains in the basal plane of the ordered, orthorhombic structure (Fig. \ref{fig:7_lps_vs_T}d) is broken by migration of O from the O1 sites to the O5 sites, leading to the tetragonal phase (Fig. \ref{fig:7_lps_vs_T}e) where these sites are randomly occupied and the chains lose long range order.
At the transition, we observe an occupancy of the O1 and O5 positions slightly higher than 0.5 atoms/f.u., where the additional O atoms come from the O4 lattice position (see Fig. \ref{fig:7_lps_vs_T}c).
With increasing temperature, the occupancy of the O1 and O5 positions steadily increases and the O4 occupation conversely decreases.

Comparison of the computational transition temperature with experiments is complicated by oxygen degassing in the experimental samples. 
In our bulk simulations, which conserve the number of particles, the orthorhombic-to-tetragonal transition is observed at lower temperatures than in experiments, consistent with the lower O1v–O5i FP formation energy of the ACE potential as compared to the DFT reference. 
However, the onset of basal-plane disorder observed in the present simulations at 700 K closely matches the experimental onset of oxygen loss \cite{expTransition}, indicating that the ACE potential accurately captures the temperatures at which oxygen diffusion becomes significant.
This suggests that increased oxygen mobility below the transition might initially deplete more effectively the newly occupied O5 sites, maintaining the orthorhombic structure up to higher temperatures as compared to our model with conserved number of particles, explaining part of the present underestimation of transition temperature.

In contrast to the more complex comparison with the experimental transition temperature, the difference observed in the $c$ lattice parameter between theory and experiment can be unambiguously attributed to oxygen degassing.
The sharper rise of $c$ observed experimentally in the transition region (Fig. \ref{fig:7_lps_vs_T}b) is due to the known increase of $c$ with decreasing oxygen content \cite{expTransition_vs_O_content}, not accounted for in our simulations.
A complete investigation of the interplay between oxygen stoichiometry and the structural transition would require a potential capable of handling variable oxygen content and simulations in the grand-canonical ensemble or with open surfaces, which is beyond the scope of this work.

\begin{figure}
    \centering
    \includegraphics[width=0.7\columnwidth]{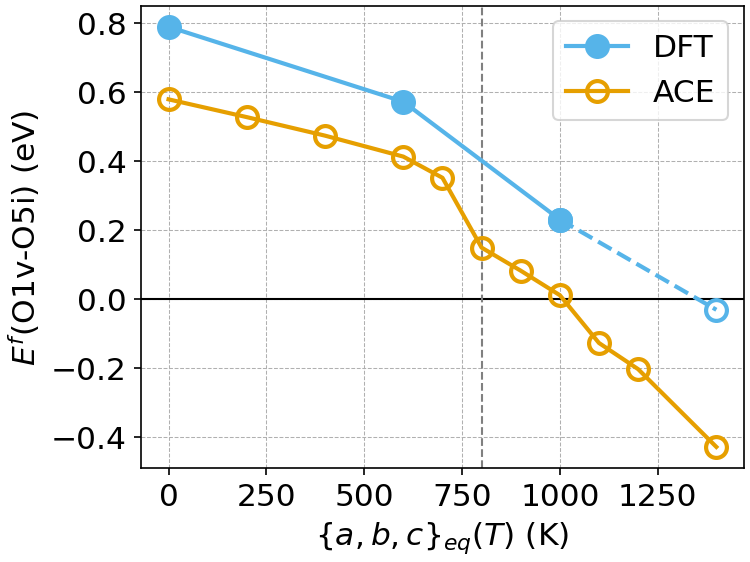}
    \caption{Formation energy of the O1v-O5i FP at the expanded lattice parameters taken from the NPT simulations ($\{a,b,c\}_{eq}(T)$ indicates the equilibrium lattice vectors at temperature T). For simplicity, the corresponding temperature is shown on the x axis. The vertical dashed line indicates the thermodynamic transition temperature. The last DFT point at 1400 K is reported as dashed because the structure with a single FP is unstable (see text).}
    \label{fig:9_Ef_O1v-O5i_vs_T}
\end{figure}

To elucidate the thermodynamic origin of the order-disorder transition, we calculate the formation energy of the O1v-O5i FP at 0 K using the expanded lattice parameters obtained from NPT simulations and relaxing only atomic positions (see Fig. \ref{fig:9_Ef_O1v-O5i_vs_T}).
The formation energy decreases with expanding lattice parameters (i.e., increasing temperature), finding that it is energetically favorable to generate at least one FP for lattice parameters corresponding to temperatures of 1000 K or higher.
Importantly, around 800 K the system is already tetragonal but the formation energy is positive, indicating that the formation of FPs is driven by entropic effects around the transition temperature rather than purely energetic ones.
To verify the reliability of the ACE potential, we recalculated the formation energy with DFT at 3 different expanded lattices corresponding to 600, 1000, and 1400 K and observed a similar decrease in formation energy.
The data point at 1400 K in Fig. \ref{fig:9_Ef_O1v-O5i_vs_T} should be taken only as an indication of the trend in the formation energy, since the structure with one FP at this expanded lattice is not stable in DFT, with the interstitial atom inducing a heavy deformation of the neighboring CuO chains.

\section{Conclusions} \label{sec:discussion}

In this work, we have developed an ACE potential for YBCO capable of describing the phase transition from the low-temperature orthorhombic structure to the high-temperature tetragonal structure.
The potential has been validated against the equation of state, lattice parameters far from equilibrium, and O FPs energetics.

We identify the mechanism driving the structural transition as the rearrangement of O atoms in the basal plane: O atoms from the CuO chains (O1) and, to a lesser extent, from the apical positions (O4) migrate to the interstitial O5 positions between the chains, effectively rendering the $a$ and $b$ directions equivalent.
The transition is driven by entropic effects that favor disordering of the O occupations in the basal plane, but it is facilitated by a reduced formation energy of O FPs at expanded lattices.

The theoretical transition temperature is in reasonable agreement with experiments (800 K and approximately 1000 K, respectively), particularly considering that heating of YBCO is accompanied by oxygen loss, which effectively changes the chemistry of the system.
Further progress will require extending the ACE potential to variable oxygen contents and employing advanced simulation techniques to access experimentally relevant timescales.

Overall, the developed potential enables the investigation of mass transport in this complex cuprate system with unprecedented accuracy. 
Beyond advancing our fundamental understanding of oxygen diffusion in HTS, this potential may also contribute to practical applications, such as optimizing annealing strategies for restoring superconductivity in radiation-damaged HTS tapes \cite{Unterrainer_annealing}.

\section{Methods} \label{sec:methods}

\subsection{Dataset creation} \label{ssec:dataset}
The data set for defect free structure, structures with single interstitials of each species, and structures with vacancies at each symmetrically inequivalent crystallographic position was created taking snapshot from MD simulations carried out with a preliminary version of the potential. 
These simulations were run with 4x4x2 supercells (416 atoms) at temperatures of 20, 250, 500, 750, 1000, and 1250 K.

To sample disordered environments, we included amorphous-like structures generated by randomly displacing the atoms in an ideal $2 \times 2 \times 1$ supercell according to a Gaussian distribution (0 - 8 \AA), and wrapping the atoms back into the box.
To avoid extreme overlap of atomic positions, these configurations were pre-relaxed for 10 steps with the purely repulsive Ziegler–Biersack–Littmark (ZBL) potential \cite{ZBL}. 
Subsequently, atomic positions were relaxed according to DFT forces for 10 ionic steps to reduce excessively high forces.

The O FPs configurations included in the data set comprised of vacancies of each inequivalent O position and interstitial O atom at the O5 position (see Fig. \ref{fig:1_combo}).
The approximate TSs were generated as the mid point between the defect free structure (initial) and the O1v-O5i, O2v-O5i, O4v-O5i FPs (final states), and between the O4v-O5i FP (initial state) and the O2v-O5i,  O3v-O5i FPs (final states).
Random perturbations of atomic positions and lattice vectors were applied to increase configurational diversity.
These FPs and TSs were selected based on physical intuition of the most likely diffusion paths for O atoms.
Input structures were modified with the atomistic simulation environment (ASE) \cite{ASE} and Atomsk \cite{Atomsk}.

\subsection{Computational details of the DFT training set and validation calculations} \label{ssec:compdetDFT}

All DFT calculations were carried out with the Vienna Ab Initio Simulation Package (VASP) \cite{VASP_I,VASP_II,VASP_III} with the projector-augmented wave (PAW) method \cite{PAW_Blochl,PAW_vasp} and the generalized-gradient approximation of Perdew, Burke and Ernzerhof (PBE) \cite{PBE}. 
Spin polarization was neglected.
The plane-waves energy cutoff was set to 600 eV, while sampling the Brillouin zone with a Gamma-centered Monkhorst-Pack \cite{MPscheme} $2 \times 2 \times 1$ k-mesh, corresponding to k-spacings below 0.09 \AA$^{-1}$ in all cases. 
These computational parameters ensure an error below 4 meV/atom compared to a calculation with an 800 eV energy cutoff and a $4 \times 4 \times 2$ k-mesh, while keeping the computational effort for dataset construction feasible.
Convergence thresholds were set to $10^{-5}$ eV in energy for the self-consistent electronic minimization, and to 0.01 eV/\AA \; on forces for structural relaxations.
Gaussian smearing of 0.05 eV was applied to the electronic states to improve convergence.

Single-point calculations were carried out for configurations with atomic displacements in the defect free case and for configurations with vacancies, interstitials, FPs and TSs. 
Defect free structures over a wide volume range were relaxed under fixed volume. 
Minimum energy paths for formation of O FPs were calculated with the climbing image Nudged Elastic Bands (NEB) method using 5 intermediate configurations along the diffusion path, with the same energy and force thresholds.

\subsection{Parameters of the ACE potential}\label{ssec:paramACE}
The final potential consists of 1000 functions per element, with a 7 \AA \; cutoff radius for interaction with neighbors.
We employed default parameters as implemented in pacemaker \cite{Pacemaker_NPJ,Pacemaker_PRM}, using simplified Bessel radial functions, the Finnis-Sinclair embedding function, and applying an energy-based weighting policy on the training set, disregarding configurations with forces larger than 25 eV/\AA.
We also included the automatic core repulsion based on the ZBL potential \cite{ZBL}.
All energies are shifted with reference to the isolated atoms, which were calculated with non-spin polarized DFT using a large box (edges of 30 \AA).
We tested alternative radial functions (exponentially-scaled Chebyshev polynomials) and increasing the number of functions to 1400 per element but observed no improvement in RMSEs or equation of state results.
The chosen potential parameters therefore ensure accuracy and computational efficiency.

\subsection{Computational details of calculations and molecular dynamics simulations with interatomic potentials}\label{ssec:compdetMD}

All calculations with the ACE and B+C potentials were carried out using LAMMPS \cite{lammps}. 
Structural relaxations for equation of state and NEB calculations were carried out using a threshold for convergence on forces of 0.01 eV/\AA.

The equation of state for the ACE potential was calculated using ASE \cite{ASE} relaxing atomic positions and lattice vectors at each volume, whereas for the B+C potential we employed LAMMPS and carried out structural relaxations at finite pressures.
The unit cell was used for these calculations.

The NPT simulations employed a Nos\'e-Hoover thermostat and barostat with damping times of 0.05 and 1 ps, respectively, using a 1 fs timestep.
All simulations employed the 0 K relaxed orthorhombic structure as initial configuration. 
Simulated time ranged between 1.6 ns and 75 ns, depending on convergence of the $a$ and $b$ lattice parameters (see Supplementary Material).
Lattice parameters were averaged over the last 1.5 ns of each simulation with values recorded every 10 fs.
Analysis of O sites occupation was carried out with OVITO \cite{ovito}.

\section*{Acknowledgements} \label{sec:acknowledgements}
The computations were enabled by resources provided by the National Academic Infrastructure for Supercomputing in Sweden (NAISS) at the National Supercomputer Centre (NSC), Linköping University, and at the PDC Center for High Performance Computing, KTH Royal Institute of Technology, partially funded by the Swedish Research Council through Grant Agreement No. 2022-06725.

DG acknowledges financial support from the Swedish Research Council (VR) through Grant No. 2023-00208.
NDE acknowledges that this publication is part of the project PNRR-NGEU which has received funding from the MUR – DM 117/2023 (or DM 118/2023). NDE, DT and FL acknowledge support from Eni S.p.A..
JB acknowledges funding from the Research council of Finland through the OCRAMLIP project, grant number 354234.
FD acknowledges the Research Council of Finland project SPATEC (Grant No 349690) for financial support.
\section*{Contributions}

D.G., D.T. and F.L. conceived the project.
D.G. and N.DE. carried out all the simulations.
J.B. assisted with the creation of the data set and the training of the potential.
J.K. and F.D. contributed to the identification of the simulation methodology.
All authors contributed to the interpretation of the results and the writing of the manuscript.


\begin{thebibliography}{77}%
\makeatletter
\providecommand \@ifxundefined [1]{%
 \@ifx{#1\undefined}
}%
\providecommand \@ifnum [1]{%
 \ifnum #1\expandafter \@firstoftwo
 \else \expandafter \@secondoftwo
 \fi
}%
\providecommand \@ifx [1]{%
 \ifx #1\expandafter \@firstoftwo
 \else \expandafter \@secondoftwo
 \fi
}%
\providecommand \natexlab [1]{#1}%
\providecommand \enquote  [1]{``#1''}%
\providecommand \bibnamefont  [1]{#1}%
\providecommand \bibfnamefont [1]{#1}%
\providecommand \citenamefont [1]{#1}%
\providecommand \href@noop [0]{\@secondoftwo}%
\providecommand \href [0]{\begingroup \@sanitize@url \@href}%
\providecommand \@href[1]{\@@startlink{#1}\@@href}%
\providecommand \@@href[1]{\endgroup#1\@@endlink}%
\providecommand \@sanitize@url [0]{\catcode `\\12\catcode `\$12\catcode `\&12\catcode `\#12\catcode `\^12\catcode `\_12\catcode `\%12\relax}%
\providecommand \@@startlink[1]{}%
\providecommand \@@endlink[0]{}%
\providecommand \url  [0]{\begingroup\@sanitize@url \@url }%
\providecommand \@url [1]{\endgroup\@href {#1}{\urlprefix }}%
\providecommand \urlprefix  [0]{URL }%
\providecommand \Eprint [0]{\href }%
\providecommand \doibase [0]{http://dx.doi.org/}%
\providecommand \selectlanguage [0]{\@gobble}%
\providecommand \bibinfo  [0]{\@secondoftwo}%
\providecommand \bibfield  [0]{\@secondoftwo}%
\providecommand \translation [1]{[#1]}%
\providecommand \BibitemOpen [0]{}%
\providecommand \bibitemStop [0]{}%
\providecommand \bibitemNoStop [0]{.\EOS\space}%
\providecommand \EOS [0]{\spacefactor3000\relax}%
\providecommand \BibitemShut  [1]{\csname bibitem#1\endcsname}%
\let\auto@bib@innerbib\@empty
\bibitem [{\citenamefont {Coombs}\ \emph {et~al.}(2024)\citenamefont {Coombs}, \citenamefont {Wang}, \citenamefont {Shah}, \citenamefont {Hu}, \citenamefont {Hao}, \citenamefont {Patel}, \citenamefont {Wei}, \citenamefont {Wu}, \citenamefont {Coombs},\ and\ \citenamefont {Wang}}]{HTS_applications}%
  \BibitemOpen
  \bibfield  {author} {\bibinfo {author} {\bibfnamefont {T.~A.}\ \bibnamefont {Coombs}}, \bibinfo {author} {\bibfnamefont {Q.}~\bibnamefont {Wang}}, \bibinfo {author} {\bibfnamefont {A.}~\bibnamefont {Shah}}, \bibinfo {author} {\bibfnamefont {J.}~\bibnamefont {Hu}}, \bibinfo {author} {\bibfnamefont {L.}~\bibnamefont {Hao}}, \bibinfo {author} {\bibfnamefont {I.}~\bibnamefont {Patel}}, \bibinfo {author} {\bibfnamefont {H.}~\bibnamefont {Wei}}, \bibinfo {author} {\bibfnamefont {Y.}~\bibnamefont {Wu}}, \bibinfo {author} {\bibfnamefont {T.}~\bibnamefont {Coombs}}, \ and\ \bibinfo {author} {\bibfnamefont {W.}~\bibnamefont {Wang}},\ }\href {\doibase 10.1038/s44287-024-00112-y} {\bibfield  {journal} {\bibinfo  {journal} {Nature Reviews Electrical Engineering}\ }\textbf {\bibinfo {volume} {1}},\ \bibinfo {pages} {788} (\bibinfo {year} {2024})}\BibitemShut {NoStop}%
\bibitem [{\citenamefont {Obradors}\ and\ \citenamefont {Puig}(2014)}]{2gHTS_2014}%
  \BibitemOpen
  \bibfield  {author} {\bibinfo {author} {\bibfnamefont {X.}~\bibnamefont {Obradors}}\ and\ \bibinfo {author} {\bibfnamefont {T.}~\bibnamefont {Puig}},\ }\href {\doibase 10.1088/0953-2048/27/4/044003} {\bibfield  {journal} {\bibinfo  {journal} {Superconductor Science and Technology}\ }\textbf {\bibinfo {volume} {27}},\ \bibinfo {pages} {044003} (\bibinfo {year} {2014})}\BibitemShut {NoStop}%
\bibitem [{\citenamefont {MacManus-Driscoll}\ and\ \citenamefont {Wimbush}(2021)}]{2gHTS_2021}%
  \BibitemOpen
  \bibfield  {author} {\bibinfo {author} {\bibfnamefont {J.~L.}\ \bibnamefont {MacManus-Driscoll}}\ and\ \bibinfo {author} {\bibfnamefont {S.~C.}\ \bibnamefont {Wimbush}},\ }\href {\doibase 10.1038/s41578-021-00290-3} {\bibfield  {journal} {\bibinfo  {journal} {Nature Reviews Materials}\ }\textbf {\bibinfo {volume} {6}},\ \bibinfo {pages} {587} (\bibinfo {year} {2021})}\BibitemShut {NoStop}%
\bibitem [{\citenamefont {Sorbom}\ \emph {et~al.}(2015)\citenamefont {Sorbom}, \citenamefont {Ball}, \citenamefont {Palmer}, \citenamefont {Mangiarotti}, \citenamefont {Sierchio}, \citenamefont {Bonoli}, \citenamefont {Kasten}, \citenamefont {Sutherland}, \citenamefont {Barnard}, \citenamefont {Haakonsen}, \citenamefont {Goh}, \citenamefont {Sung},\ and\ \citenamefont {Whyte}}]{ARC_original}%
  \BibitemOpen
  \bibfield  {author} {\bibinfo {author} {\bibfnamefont {B.~N.}\ \bibnamefont {Sorbom}}, \bibinfo {author} {\bibfnamefont {J.}~\bibnamefont {Ball}}, \bibinfo {author} {\bibfnamefont {T.~R.}\ \bibnamefont {Palmer}}, \bibinfo {author} {\bibfnamefont {F.~J.}\ \bibnamefont {Mangiarotti}}, \bibinfo {author} {\bibfnamefont {J.~M.}\ \bibnamefont {Sierchio}}, \bibinfo {author} {\bibfnamefont {P.}~\bibnamefont {Bonoli}}, \bibinfo {author} {\bibfnamefont {C.}~\bibnamefont {Kasten}}, \bibinfo {author} {\bibfnamefont {D.~A.}\ \bibnamefont {Sutherland}}, \bibinfo {author} {\bibfnamefont {H.~S.}\ \bibnamefont {Barnard}}, \bibinfo {author} {\bibfnamefont {C.~B.}\ \bibnamefont {Haakonsen}}, \bibinfo {author} {\bibfnamefont {J.}~\bibnamefont {Goh}}, \bibinfo {author} {\bibfnamefont {C.}~\bibnamefont {Sung}}, \ and\ \bibinfo {author} {\bibfnamefont {D.~G.}\ \bibnamefont {Whyte}},\ }\href {\doibase 10.1016/j.fusengdes.2015.07.008} {\bibfield  {journal} {\bibinfo  {journal} {Fusion Engineering and Design}\ }\textbf {\bibinfo
  {volume} {100}},\ \bibinfo {pages} {378} (\bibinfo {year} {2015})}\BibitemShut {NoStop}%
\bibitem [{\citenamefont {Vieira}\ \emph {et~al.}(2024)\citenamefont {Vieira}, \citenamefont {Arsenault}, \citenamefont {Barnett}, \citenamefont {Bartoszek}, \citenamefont {Beck}, \citenamefont {Chamberlain}, \citenamefont {Cheng}, \citenamefont {Dombrowski}, \citenamefont {Doody}, \citenamefont {Dunn}, \citenamefont {Estrada}, \citenamefont {Fry}, \citenamefont {Garberg}, \citenamefont {Golfinopoulos}, \citenamefont {Greenberg}, \citenamefont {Heller}, \citenamefont {Hubbard}, \citenamefont {Korsun}, \citenamefont {Kuznetsov}, \citenamefont {LaBombard}, \citenamefont {Lammi}, \citenamefont {Leccacorvi}, \citenamefont {Levine}, \citenamefont {Mendoza}, \citenamefont {Metcalfe}, \citenamefont {Michael}, \citenamefont {Mouratidis}, \citenamefont {Mumgaard}, \citenamefont {Muncks}, \citenamefont {Murray}, \citenamefont {Nash}, \citenamefont {Pfeiffer}, \citenamefont {Pierson}, \citenamefont {Radovinsky}, \citenamefont {Rosati}, \citenamefont {Rowell}, \citenamefont {Salazar}, \citenamefont {Schweiger},
  \citenamefont {Shiraiwa}, \citenamefont {Sorbom}, \citenamefont {Stahle}, \citenamefont {Stevens}, \citenamefont {Tammana}, \citenamefont {Toland}, \citenamefont {Vernacchia}, \citenamefont {Voirin}, \citenamefont {Warner}, \citenamefont {Watterson}, \citenamefont {Whyte}, \citenamefont {Wilcox}, \citenamefont {Zhou}, \citenamefont {Zhukovsky},\ and\ \citenamefont {Hartwig}}]{SPARC_magnet}%
  \BibitemOpen
  \bibfield  {author} {\bibinfo {author} {\bibfnamefont {R.~F.}\ \bibnamefont {Vieira}}, \bibinfo {author} {\bibfnamefont {D.}~\bibnamefont {Arsenault}}, \bibinfo {author} {\bibfnamefont {R.}~\bibnamefont {Barnett}}, \bibinfo {author} {\bibfnamefont {L.}~\bibnamefont {Bartoszek}}, \bibinfo {author} {\bibfnamefont {W.}~\bibnamefont {Beck}}, \bibinfo {author} {\bibfnamefont {S.}~\bibnamefont {Chamberlain}}, \bibinfo {author} {\bibfnamefont {J.~L.}\ \bibnamefont {Cheng}}, \bibinfo {author} {\bibfnamefont {E.}~\bibnamefont {Dombrowski}}, \bibinfo {author} {\bibfnamefont {J.}~\bibnamefont {Doody}}, \bibinfo {author} {\bibfnamefont {D.}~\bibnamefont {Dunn}}, \bibinfo {author} {\bibfnamefont {J.}~\bibnamefont {Estrada}}, \bibinfo {author} {\bibfnamefont {V.}~\bibnamefont {Fry}}, \bibinfo {author} {\bibfnamefont {S.}~\bibnamefont {Garberg}}, \bibinfo {author} {\bibfnamefont {T.}~\bibnamefont {Golfinopoulos}}, \bibinfo {author} {\bibfnamefont {A.}~\bibnamefont {Greenberg}}, \bibinfo {author} {\bibfnamefont
  {S.}~\bibnamefont {Heller}}, \bibinfo {author} {\bibfnamefont {A.}~\bibnamefont {Hubbard}}, \bibinfo {author} {\bibfnamefont {D.}~\bibnamefont {Korsun}}, \bibinfo {author} {\bibfnamefont {S.}~\bibnamefont {Kuznetsov}}, \bibinfo {author} {\bibfnamefont {B.}~\bibnamefont {LaBombard}}, \bibinfo {author} {\bibfnamefont {C.}~\bibnamefont {Lammi}}, \bibinfo {author} {\bibfnamefont {R.}~\bibnamefont {Leccacorvi}}, \bibinfo {author} {\bibfnamefont {M.}~\bibnamefont {Levine}}, \bibinfo {author} {\bibfnamefont {D.~C.}\ \bibnamefont {Mendoza}}, \bibinfo {author} {\bibfnamefont {K.}~\bibnamefont {Metcalfe}}, \bibinfo {author} {\bibfnamefont {P.}~\bibnamefont {Michael}}, \bibinfo {author} {\bibfnamefont {T.}~\bibnamefont {Mouratidis}}, \bibinfo {author} {\bibfnamefont {R.}~\bibnamefont {Mumgaard}}, \bibinfo {author} {\bibfnamefont {J.~P.}\ \bibnamefont {Muncks}}, \bibinfo {author} {\bibfnamefont {R.}~\bibnamefont {Murray}}, \bibinfo {author} {\bibfnamefont {D.}~\bibnamefont {Nash}}, \bibinfo {author} {\bibfnamefont
  {A.}~\bibnamefont {Pfeiffer}}, \bibinfo {author} {\bibfnamefont {S.}~\bibnamefont {Pierson}}, \bibinfo {author} {\bibfnamefont {A.}~\bibnamefont {Radovinsky}}, \bibinfo {author} {\bibfnamefont {R.}~\bibnamefont {Rosati}}, \bibinfo {author} {\bibfnamefont {M.}~\bibnamefont {Rowell}}, \bibinfo {author} {\bibfnamefont {E.}~\bibnamefont {Salazar}}, \bibinfo {author} {\bibfnamefont {S.}~\bibnamefont {Schweiger}}, \bibinfo {author} {\bibfnamefont {S.}~\bibnamefont {Shiraiwa}}, \bibinfo {author} {\bibfnamefont {B.}~\bibnamefont {Sorbom}}, \bibinfo {author} {\bibfnamefont {P.}~\bibnamefont {Stahle}}, \bibinfo {author} {\bibfnamefont {K.}~\bibnamefont {Stevens}}, \bibinfo {author} {\bibfnamefont {D.}~\bibnamefont {Tammana}}, \bibinfo {author} {\bibfnamefont {T.}~\bibnamefont {Toland}}, \bibinfo {author} {\bibfnamefont {M.}~\bibnamefont {Vernacchia}}, \bibinfo {author} {\bibfnamefont {E.}~\bibnamefont {Voirin}}, \bibinfo {author} {\bibfnamefont {A.}~\bibnamefont {Warner}}, \bibinfo {author} {\bibfnamefont
  {A.}~\bibnamefont {Watterson}}, \bibinfo {author} {\bibfnamefont {D.~G.}\ \bibnamefont {Whyte}}, \bibinfo {author} {\bibfnamefont {S.}~\bibnamefont {Wilcox}}, \bibinfo {author} {\bibfnamefont {L.}~\bibnamefont {Zhou}}, \bibinfo {author} {\bibfnamefont {A.}~\bibnamefont {Zhukovsky}}, \ and\ \bibinfo {author} {\bibfnamefont {Z.~S.}\ \bibnamefont {Hartwig}},\ }\href {\doibase 10.1109/TASC.2024.3356571} {\bibfield  {journal} {\bibinfo  {journal} {IEEE Transactions on Applied Superconductivity}\ }\textbf {\bibinfo {volume} {34}},\ \bibinfo {pages} {1} (\bibinfo {year} {2024})}\BibitemShut {NoStop}%
\bibitem [{\citenamefont {Kingham}\ and\ \citenamefont {Gryaznevich}(2024)}]{TE}%
  \BibitemOpen
  \bibfield  {author} {\bibinfo {author} {\bibfnamefont {D.}~\bibnamefont {Kingham}}\ and\ \bibinfo {author} {\bibfnamefont {M.}~\bibnamefont {Gryaznevich}},\ }\href {\doibase 10.1063/5.0170088} {\bibfield  {journal} {\bibinfo  {journal} {Physics of Plasmas}\ }\textbf {\bibinfo {volume} {31}},\ \bibinfo {pages} {042507} (\bibinfo {year} {2024})}\BibitemShut {NoStop}%
\bibitem [{\citenamefont {Meschini}\ \emph {et~al.}(2023)\citenamefont {Meschini}, \citenamefont {Laviano}, \citenamefont {Ledda}, \citenamefont {Pettinari}, \citenamefont {Testoni}, \citenamefont {Torsello},\ and\ \citenamefont {Panella}}]{Commercial_fusion}%
  \BibitemOpen
  \bibfield  {author} {\bibinfo {author} {\bibfnamefont {S.}~\bibnamefont {Meschini}}, \bibinfo {author} {\bibfnamefont {F.}~\bibnamefont {Laviano}}, \bibinfo {author} {\bibfnamefont {F.}~\bibnamefont {Ledda}}, \bibinfo {author} {\bibfnamefont {D.}~\bibnamefont {Pettinari}}, \bibinfo {author} {\bibfnamefont {R.}~\bibnamefont {Testoni}}, \bibinfo {author} {\bibfnamefont {D.}~\bibnamefont {Torsello}}, \ and\ \bibinfo {author} {\bibfnamefont {B.}~\bibnamefont {Panella}},\ }\href {https://www.frontiersin.org/journals/energy-research/articles/10.3389/fenrg.2023.1157394/full} {\bibfield  {journal} {\bibinfo  {journal} {Frontiers in Energy Research}\ }\textbf {\bibinfo {volume} {11}} (\bibinfo {year} {2023})}\BibitemShut {NoStop}%
\bibitem [{\citenamefont {Li}\ \emph {et~al.}(2024)\citenamefont {Li}, \citenamefont {Pan}, \citenamefont {Zhang}, \citenamefont {Zhu}, \citenamefont {Zhang}, \citenamefont {Zhang}, \citenamefont {Dong}, \citenamefont {Ye},\ and\ \citenamefont {Yang}}]{HTS_tokamak_2024}%
  \BibitemOpen
  \bibfield  {author} {\bibinfo {author} {\bibfnamefont {Z.~Y.}\ \bibnamefont {Li}}, \bibinfo {author} {\bibfnamefont {Z.~C.}\ \bibnamefont {Pan}}, \bibinfo {author} {\bibfnamefont {Q.~J.}\ \bibnamefont {Zhang}}, \bibinfo {author} {\bibfnamefont {K.~P.}\ \bibnamefont {Zhu}}, \bibinfo {author} {\bibfnamefont {C.}~\bibnamefont {Zhang}}, \bibinfo {author} {\bibfnamefont {Z.~W.}\ \bibnamefont {Zhang}}, \bibinfo {author} {\bibfnamefont {G.}~\bibnamefont {Dong}}, \bibinfo {author} {\bibfnamefont {Y.~M.}\ \bibnamefont {Ye}}, \ and\ \bibinfo {author} {\bibfnamefont {Z.}~\bibnamefont {Yang}},\ }\href {\doibase 10.1016/j.supcon.2024.100137} {\bibfield  {journal} {\bibinfo  {journal} {Superconductivity}\ }\textbf {\bibinfo {volume} {12}},\ \bibinfo {pages} {100137} (\bibinfo {year} {2024})}\BibitemShut {NoStop}%
\bibitem [{\citenamefont {Sarasola}\ \emph {et~al.}(2020)\citenamefont {Sarasola}, \citenamefont {Wesche}, \citenamefont {Ivashov}, \citenamefont {Sedlak}, \citenamefont {Uglietti},\ and\ \citenamefont {Bruzzone}}]{DEMO_CS}%
  \BibitemOpen
  \bibfield  {author} {\bibinfo {author} {\bibfnamefont {X.}~\bibnamefont {Sarasola}}, \bibinfo {author} {\bibfnamefont {R.}~\bibnamefont {Wesche}}, \bibinfo {author} {\bibfnamefont {I.}~\bibnamefont {Ivashov}}, \bibinfo {author} {\bibfnamefont {K.}~\bibnamefont {Sedlak}}, \bibinfo {author} {\bibfnamefont {D.}~\bibnamefont {Uglietti}}, \ and\ \bibinfo {author} {\bibfnamefont {P.}~\bibnamefont {Bruzzone}},\ }\href {\doibase 10.1109/TASC.2020.2965066} {\bibfield  {journal} {\bibinfo  {journal} {IEEE Transactions on Applied Superconductivity}\ }\textbf {\bibinfo {volume} {30}},\ \bibinfo {pages} {1} (\bibinfo {year} {2020})}\BibitemShut {NoStop}%
\bibitem [{\citenamefont {Federici}\ \emph {et~al.}(2024)\citenamefont {Federici}, \citenamefont {Siccinio}, \citenamefont {Bachmann}, \citenamefont {Giannini}, \citenamefont {Luongo},\ and\ \citenamefont {Lungaroni}}]{DEMO_HTS}%
  \BibitemOpen
  \bibfield  {author} {\bibinfo {author} {\bibfnamefont {G.}~\bibnamefont {Federici}}, \bibinfo {author} {\bibfnamefont {M.}~\bibnamefont {Siccinio}}, \bibinfo {author} {\bibfnamefont {C.}~\bibnamefont {Bachmann}}, \bibinfo {author} {\bibfnamefont {L.}~\bibnamefont {Giannini}}, \bibinfo {author} {\bibfnamefont {C.}~\bibnamefont {Luongo}}, \ and\ \bibinfo {author} {\bibfnamefont {M.}~\bibnamefont {Lungaroni}},\ }\href {\doibase 10.1088/1741-4326/ad2425} {\bibfield  {journal} {\bibinfo  {journal} {Nuclear Fusion}\ }\textbf {\bibinfo {volume} {64}},\ \bibinfo {pages} {036025} (\bibinfo {year} {2024})}\BibitemShut {NoStop}%
\bibitem [{\citenamefont {Bottura}\ \emph {et~al.}(2024)\citenamefont {Bottura}, \citenamefont {Accettura}, \citenamefont {Amemiya}, \citenamefont {Auchmann}, \citenamefont {Berg}, \citenamefont {Bersani}, \citenamefont {Bertarelli}, \citenamefont {Boattini}, \citenamefont {Bordini}, \citenamefont {Borges~de Sousa}, \citenamefont {Breschi}, \citenamefont {Caiffi}, \citenamefont {Chaud}, \citenamefont {Debray}, \citenamefont {Dudarev}, \citenamefont {Eisterer}, \citenamefont {Fabbri}, \citenamefont {Farinon}, \citenamefont {Ferracin}, \citenamefont {De~Gersem}, \citenamefont {Kario}, \citenamefont {Kolehmainen}, \citenamefont {Kosse}, \citenamefont {Lorenzo~Gomez}, \citenamefont {Losito}, \citenamefont {Mariotto}, \citenamefont {Mentink}, \citenamefont {Mulder}, \citenamefont {Musenich}, \citenamefont {Novelli}, \citenamefont {Ogitsu}, \citenamefont {Palmer}, \citenamefont {Pavan}, \citenamefont {Piekarz}, \citenamefont {Portone}, \citenamefont {Quettier}, \citenamefont {Rochepault}, \citenamefont {Rossi},
  \citenamefont {Salmi}, \citenamefont {Schneider-Muntau}, \citenamefont {Senatore}, \citenamefont {Statera}, \citenamefont {Ten~Kate}, \citenamefont {Testoni}, \citenamefont {Vallone}, \citenamefont {Verweij}, \citenamefont {Van~Weelderen}, \citenamefont {Wozniak}, \citenamefont {Yamamoto}, \citenamefont {Yang}, \citenamefont {Zhai},\ and\ \citenamefont {Zlobin}}]{Muon_colliders}%
  \BibitemOpen
  \bibfield  {author} {\bibinfo {author} {\bibfnamefont {L.}~\bibnamefont {Bottura}}, \bibinfo {author} {\bibfnamefont {C.}~\bibnamefont {Accettura}}, \bibinfo {author} {\bibfnamefont {N.}~\bibnamefont {Amemiya}}, \bibinfo {author} {\bibfnamefont {B.}~\bibnamefont {Auchmann}}, \bibinfo {author} {\bibfnamefont {J.}~\bibnamefont {Berg}}, \bibinfo {author} {\bibfnamefont {A.}~\bibnamefont {Bersani}}, \bibinfo {author} {\bibfnamefont {A.}~\bibnamefont {Bertarelli}}, \bibinfo {author} {\bibfnamefont {F.}~\bibnamefont {Boattini}}, \bibinfo {author} {\bibfnamefont {B.}~\bibnamefont {Bordini}}, \bibinfo {author} {\bibfnamefont {P.}~\bibnamefont {Borges~de Sousa}}, \bibinfo {author} {\bibfnamefont {M.}~\bibnamefont {Breschi}}, \bibinfo {author} {\bibfnamefont {B.}~\bibnamefont {Caiffi}}, \bibinfo {author} {\bibfnamefont {X.}~\bibnamefont {Chaud}}, \bibinfo {author} {\bibfnamefont {F.}~\bibnamefont {Debray}}, \bibinfo {author} {\bibfnamefont {A.}~\bibnamefont {Dudarev}}, \bibinfo {author} {\bibfnamefont
  {M.}~\bibnamefont {Eisterer}}, \bibinfo {author} {\bibfnamefont {S.}~\bibnamefont {Fabbri}}, \bibinfo {author} {\bibfnamefont {S.}~\bibnamefont {Farinon}}, \bibinfo {author} {\bibfnamefont {P.}~\bibnamefont {Ferracin}}, \bibinfo {author} {\bibfnamefont {H.}~\bibnamefont {De~Gersem}}, \bibinfo {author} {\bibfnamefont {A.}~\bibnamefont {Kario}}, \bibinfo {author} {\bibfnamefont {A.}~\bibnamefont {Kolehmainen}}, \bibinfo {author} {\bibfnamefont {J.}~\bibnamefont {Kosse}}, \bibinfo {author} {\bibfnamefont {J.}~\bibnamefont {Lorenzo~Gomez}}, \bibinfo {author} {\bibfnamefont {R.}~\bibnamefont {Losito}}, \bibinfo {author} {\bibfnamefont {S.}~\bibnamefont {Mariotto}}, \bibinfo {author} {\bibfnamefont {M.}~\bibnamefont {Mentink}}, \bibinfo {author} {\bibfnamefont {T.}~\bibnamefont {Mulder}}, \bibinfo {author} {\bibfnamefont {R.}~\bibnamefont {Musenich}}, \bibinfo {author} {\bibfnamefont {D.}~\bibnamefont {Novelli}}, \bibinfo {author} {\bibfnamefont {T.}~\bibnamefont {Ogitsu}}, \bibinfo {author} {\bibfnamefont
  {M.}~\bibnamefont {Palmer}}, \bibinfo {author} {\bibfnamefont {J.}~\bibnamefont {Pavan}}, \bibinfo {author} {\bibfnamefont {H.}~\bibnamefont {Piekarz}}, \bibinfo {author} {\bibfnamefont {A.}~\bibnamefont {Portone}}, \bibinfo {author} {\bibfnamefont {L.}~\bibnamefont {Quettier}}, \bibinfo {author} {\bibfnamefont {E.}~\bibnamefont {Rochepault}}, \bibinfo {author} {\bibfnamefont {L.}~\bibnamefont {Rossi}}, \bibinfo {author} {\bibfnamefont {T.}~\bibnamefont {Salmi}}, \bibinfo {author} {\bibfnamefont {H.}~\bibnamefont {Schneider-Muntau}}, \bibinfo {author} {\bibfnamefont {C.}~\bibnamefont {Senatore}}, \bibinfo {author} {\bibfnamefont {M.}~\bibnamefont {Statera}}, \bibinfo {author} {\bibfnamefont {H.}~\bibnamefont {Ten~Kate}}, \bibinfo {author} {\bibfnamefont {P.}~\bibnamefont {Testoni}}, \bibinfo {author} {\bibfnamefont {G.}~\bibnamefont {Vallone}}, \bibinfo {author} {\bibfnamefont {A.}~\bibnamefont {Verweij}}, \bibinfo {author} {\bibfnamefont {R.}~\bibnamefont {Van~Weelderen}}, \bibinfo {author} {\bibfnamefont
  {M.}~\bibnamefont {Wozniak}}, \bibinfo {author} {\bibfnamefont {A.}~\bibnamefont {Yamamoto}}, \bibinfo {author} {\bibfnamefont {Y.}~\bibnamefont {Yang}}, \bibinfo {author} {\bibfnamefont {Y.}~\bibnamefont {Zhai}}, \ and\ \bibinfo {author} {\bibfnamefont {A.}~\bibnamefont {Zlobin}},\ }\href {\doibase 10.1109/TASC.2024.3382069} {\bibfield  {journal} {\bibinfo  {journal} {IEEE Transactions on Applied Superconductivity}\ }\textbf {\bibinfo {volume} {34}},\ \bibinfo {pages} {1} (\bibinfo {year} {2024})}\BibitemShut {NoStop}%
\bibitem [{\citenamefont {Mariotto}\ \emph {et~al.}(2025)\citenamefont {Mariotto}, \citenamefont {Busatto}, \citenamefont {Calzolaio}, \citenamefont {Rossi}, \citenamefont {Sanfilippo},\ and\ \citenamefont {Sorti}}]{HTS_PSI}%
  \BibitemOpen
  \bibfield  {author} {\bibinfo {author} {\bibfnamefont {S.}~\bibnamefont {Mariotto}}, \bibinfo {author} {\bibfnamefont {S.}~\bibnamefont {Busatto}}, \bibinfo {author} {\bibfnamefont {C.}~\bibnamefont {Calzolaio}}, \bibinfo {author} {\bibfnamefont {L.}~\bibnamefont {Rossi}}, \bibinfo {author} {\bibfnamefont {S.}~\bibnamefont {Sanfilippo}}, \ and\ \bibinfo {author} {\bibfnamefont {S.}~\bibnamefont {Sorti}},\ }\href {\doibase 10.1109/TASC.2025.3543325} {\bibfield  {journal} {\bibinfo  {journal} {IEEE Transactions on Applied Superconductivity}\ }\textbf {\bibinfo {volume} {35}},\ \bibinfo {pages} {1} (\bibinfo {year} {2025})}\BibitemShut {NoStop}%
\bibitem [{\citenamefont {Giannini}\ \emph {et~al.}(2025)\citenamefont {Giannini}, \citenamefont {Luongo}, \citenamefont {Bottura}, \citenamefont {Bordini}, \citenamefont {Lechner}, \citenamefont {Kolehmainen}, \citenamefont {Leichtle}, \citenamefont {Portone}, \citenamefont {Testoni}, \citenamefont {Bajari}, \citenamefont {Siccinio}, \citenamefont {Bachmann},\ and\ \citenamefont {Federici}}]{HTS_CERN}%
  \BibitemOpen
  \bibfield  {author} {\bibinfo {author} {\bibfnamefont {L.}~\bibnamefont {Giannini}}, \bibinfo {author} {\bibfnamefont {C.}~\bibnamefont {Luongo}}, \bibinfo {author} {\bibfnamefont {L.}~\bibnamefont {Bottura}}, \bibinfo {author} {\bibfnamefont {B.}~\bibnamefont {Bordini}}, \bibinfo {author} {\bibfnamefont {A.}~\bibnamefont {Lechner}}, \bibinfo {author} {\bibfnamefont {A.}~\bibnamefont {Kolehmainen}}, \bibinfo {author} {\bibfnamefont {D.}~\bibnamefont {Leichtle}}, \bibinfo {author} {\bibfnamefont {A.}~\bibnamefont {Portone}}, \bibinfo {author} {\bibfnamefont {P.}~\bibnamefont {Testoni}}, \bibinfo {author} {\bibfnamefont {J.}~\bibnamefont {Bajari}}, \bibinfo {author} {\bibfnamefont {M.}~\bibnamefont {Siccinio}}, \bibinfo {author} {\bibfnamefont {C.}~\bibnamefont {Bachmann}}, \ and\ \bibinfo {author} {\bibfnamefont {G.}~\bibnamefont {Federici}},\ }\href {\doibase https://doi.org/10.1016/j.fusengdes.2025.114899} {\bibfield  {journal} {\bibinfo  {journal} {Fusion Engineering and Design}\ }\textbf {\bibinfo
  {volume} {214}},\ \bibinfo {pages} {114899} (\bibinfo {year} {2025})}\BibitemShut {NoStop}%
\bibitem [{\citenamefont {Bednorz}\ and\ \citenamefont {Müller}(1986)}]{HTS_discovery}%
  \BibitemOpen
  \bibfield  {author} {\bibinfo {author} {\bibfnamefont {J.~G.}\ \bibnamefont {Bednorz}}\ and\ \bibinfo {author} {\bibfnamefont {K.~A.}\ \bibnamefont {Müller}},\ }\href {\doibase 10.1007/BF01303701} {\bibfield  {journal} {\bibinfo  {journal} {Zeitschrift für Physik B Condensed Matter}\ }\textbf {\bibinfo {volume} {64}},\ \bibinfo {pages} {189} (\bibinfo {year} {1986})}\BibitemShut {NoStop}%
\bibitem [{\citenamefont {Tsuei}\ and\ \citenamefont {Kirtley}(2000)}]{Unconventional_superconductivity_I}%
  \BibitemOpen
  \bibfield  {author} {\bibinfo {author} {\bibfnamefont {C.~C.}\ \bibnamefont {Tsuei}}\ and\ \bibinfo {author} {\bibfnamefont {J.~R.}\ \bibnamefont {Kirtley}},\ }\href {\doibase 10.1103/RevModPhys.72.969} {\bibfield  {journal} {\bibinfo  {journal} {Reviews of Modern Physics}\ }\textbf {\bibinfo {volume} {72}},\ \bibinfo {pages} {969} (\bibinfo {year} {2000})}\BibitemShut {NoStop}%
\bibitem [{\citenamefont {Taillefer}(2010)}]{Unconventional_superconductivity_II}%
  \BibitemOpen
  \bibfield  {author} {\bibinfo {author} {\bibfnamefont {L.}~\bibnamefont {Taillefer}},\ }\href {\doibase 10.1146/annurev-conmatphys-070909-104117} {\bibfield  {journal} {\bibinfo  {journal} {Annual Review of Condensed Matter Physics}\ }\textbf {\bibinfo {volume} {1}},\ \bibinfo {pages} {51} (\bibinfo {year} {2010})}\BibitemShut {NoStop}%
\bibitem [{\citenamefont {Norman}(2011)}]{Unconventional_superconductivity_III}%
  \BibitemOpen
  \bibfield  {author} {\bibinfo {author} {\bibfnamefont {M.~R.}\ \bibnamefont {Norman}},\ }\href {\doibase 10.1126/science.1200181} {\bibfield  {journal} {\bibinfo  {journal} {Science}\ }\textbf {\bibinfo {volume} {332}},\ \bibinfo {pages} {196} (\bibinfo {year} {2011})}\BibitemShut {NoStop}%
\bibitem [{\citenamefont {Scalapino}(2012)}]{Unconventional_superconductivity_IV}%
  \BibitemOpen
  \bibfield  {author} {\bibinfo {author} {\bibfnamefont {D.~J.}\ \bibnamefont {Scalapino}},\ }\href {\doibase 10.1103/RevModPhys.84.1383} {\bibfield  {journal} {\bibinfo  {journal} {Reviews of Modern Physics}\ }\textbf {\bibinfo {volume} {84}},\ \bibinfo {pages} {1383} (\bibinfo {year} {2012})}\BibitemShut {NoStop}%
\bibitem [{\citenamefont {Stewart}(2017)}]{Unconventional_superconductivity_V}%
  \BibitemOpen
  \bibfield  {author} {\bibinfo {author} {\bibfnamefont {G.~R.}\ \bibnamefont {Stewart}},\ }\href {\doibase 10.1080/00018732.2017.1331615} {\bibfield  {journal} {\bibinfo  {journal} {Advances in Physics}\ }\textbf {\bibinfo {volume} {66}},\ \bibinfo {pages} {75} (\bibinfo {year} {2017})}\BibitemShut {NoStop}%
\bibitem [{\citenamefont {Tallon}(2005)}]{YBCO_phase_diagram_new}%
  \BibitemOpen
  \bibfield  {author} {\bibinfo {author} {\bibfnamefont {J.~L.}\ \bibnamefont {Tallon}},\ }in\ \href {\doibase 10.1007/3-540-27294-1_7} {\emph {\bibinfo {booktitle} {Frontiers in {Superconducting} {Materials}}}},\ \bibinfo {editor} {edited by\ \bibinfo {editor} {\bibfnamefont {A.~V.}\ \bibnamefont {Narlikar}}}\ (\bibinfo  {publisher} {Springer},\ \bibinfo {address} {Berlin, Heidelberg},\ \bibinfo {year} {2005})\ pp.\ \bibinfo {pages} {295--330}\BibitemShut {NoStop}%
\bibitem [{\citenamefont {Johnson}\ \emph {et~al.}(2008)\citenamefont {Johnson}, \citenamefont {Bording},\ and\ \citenamefont {Zhu}}]{YBCO_twinning_I}%
  \BibitemOpen
  \bibfield  {author} {\bibinfo {author} {\bibfnamefont {C.~L.}\ \bibnamefont {Johnson}}, \bibinfo {author} {\bibfnamefont {J.~K.}\ \bibnamefont {Bording}}, \ and\ \bibinfo {author} {\bibfnamefont {Y.}~\bibnamefont {Zhu}},\ }\href {\doibase 10.1103/PhysRevB.78.014517} {\bibfield  {journal} {\bibinfo  {journal} {Physical Review B}\ }\textbf {\bibinfo {volume} {78}},\ \bibinfo {pages} {014517} (\bibinfo {year} {2008})}\BibitemShut {NoStop}%
\bibitem [{\citenamefont {Borovik}\ \emph {et~al.}(2004)\citenamefont {Borovik}, \citenamefont {Malyshevsky},\ and\ \citenamefont {Rahimov}}]{YBCO_O_disorder_RBS}%
  \BibitemOpen
  \bibfield  {author} {\bibinfo {author} {\bibfnamefont {A.~S.}\ \bibnamefont {Borovik}}, \bibinfo {author} {\bibfnamefont {V.~S.}\ \bibnamefont {Malyshevsky}}, \ and\ \bibinfo {author} {\bibfnamefont {S.~V.}\ \bibnamefont {Rahimov}},\ }\href {\doibase 10.1016/j.nimb.2004.06.032} {\bibfield  {journal} {\bibinfo  {journal} {Nuclear Instruments and Methods in Physics Research Section B: Beam Interactions with Materials and Atoms}\ }\textbf {\bibinfo {volume} {226}},\ \bibinfo {pages} {385} (\bibinfo {year} {2004})}\BibitemShut {NoStop}%
\bibitem [{\citenamefont {Balatsky}\ \emph {et~al.}(2006)\citenamefont {Balatsky}, \citenamefont {Vekhter},\ and\ \citenamefont {Zhu}}]{Defect_scattering_in_superconductors}%
  \BibitemOpen
  \bibfield  {author} {\bibinfo {author} {\bibfnamefont {A.~V.}\ \bibnamefont {Balatsky}}, \bibinfo {author} {\bibfnamefont {I.}~\bibnamefont {Vekhter}}, \ and\ \bibinfo {author} {\bibfnamefont {J.-X.}\ \bibnamefont {Zhu}},\ }\href {\doibase 10.1103/RevModPhys.78.373} {\bibfield  {journal} {\bibinfo  {journal} {Rev. Mod. Phys.}\ }\textbf {\bibinfo {volume} {78}},\ \bibinfo {pages} {373} (\bibinfo {year} {2006})}\BibitemShut {NoStop}%
\bibitem [{\citenamefont {Feighan}\ \emph {et~al.}(2017)\citenamefont {Feighan}, \citenamefont {Kursumovic},\ and\ \citenamefont {MacManus-Driscoll}}]{HTS_pinning_centers_review}%
  \BibitemOpen
  \bibfield  {author} {\bibinfo {author} {\bibfnamefont {J.~P.~F.}\ \bibnamefont {Feighan}}, \bibinfo {author} {\bibfnamefont {A.}~\bibnamefont {Kursumovic}}, \ and\ \bibinfo {author} {\bibfnamefont {J.~L.}\ \bibnamefont {MacManus-Driscoll}},\ }\href {\doibase 10.1088/1361-6668/aa90d1} {\bibfield  {journal} {\bibinfo  {journal} {Superconductor Science and Technology}\ }\textbf {\bibinfo {volume} {30}},\ \bibinfo {pages} {123001} (\bibinfo {year} {2017})}\BibitemShut {NoStop}%
\bibitem [{\citenamefont {Obradors}\ and\ \citenamefont {Puig}(2024)}]{obradors_pin_2024}%
  \BibitemOpen
  \bibfield  {author} {\bibinfo {author} {\bibfnamefont {X.}~\bibnamefont {Obradors}}\ and\ \bibinfo {author} {\bibfnamefont {T.}~\bibnamefont {Puig}},\ }\href {\doibase 10.1038/s41563-024-01990-1} {\bibfield  {journal} {\bibinfo  {journal} {Nature Materials}\ }\textbf {\bibinfo {volume} {23}},\ \bibinfo {pages} {1311} (\bibinfo {year} {2024})}\BibitemShut {NoStop}%
\bibitem [{\citenamefont {Puig}\ \emph {et~al.}(2024)\citenamefont {Puig}, \citenamefont {Gutierrez},\ and\ \citenamefont {Obradors}}]{puig_impact_2024}%
  \BibitemOpen
  \bibfield  {author} {\bibinfo {author} {\bibfnamefont {T.}~\bibnamefont {Puig}}, \bibinfo {author} {\bibfnamefont {J.}~\bibnamefont {Gutierrez}}, \ and\ \bibinfo {author} {\bibfnamefont {X.}~\bibnamefont {Obradors}},\ }\href {\doibase 10.1038/s42254-023-00663-3} {\bibfield  {journal} {\bibinfo  {journal} {Nature Reviews Physics}\ }\textbf {\bibinfo {volume} {6}},\ \bibinfo {pages} {132} (\bibinfo {year} {2024})}\BibitemShut {NoStop}%
\bibitem [{\citenamefont {Ruiz}\ \emph {et~al.}(2026)\citenamefont {Ruiz}, \citenamefont {Hänisch}, \citenamefont {Polichetti}, \citenamefont {Galluzzi}, \citenamefont {Gozzelino}, \citenamefont {Torsello}, \citenamefont {Milošević-Govedarović}, \citenamefont {Grbović-Novaković}, \citenamefont {Dobrovolskiy}, \citenamefont {Lang}, \citenamefont {Grimaldi}, \citenamefont {Crisan}, \citenamefont {Ionescu}, \citenamefont {Cayado}, \citenamefont {Willa}, \citenamefont {Barbiellini}, \citenamefont {Eley},\ and\ \citenamefont {Badía-Majós}}]{ruiz_critical_2026}%
  \BibitemOpen
  \bibfield  {author} {\bibinfo {author} {\bibfnamefont {H.~S.}\ \bibnamefont {Ruiz}}, \bibinfo {author} {\bibfnamefont {J.}~\bibnamefont {Hänisch}}, \bibinfo {author} {\bibfnamefont {M.}~\bibnamefont {Polichetti}}, \bibinfo {author} {\bibfnamefont {A.}~\bibnamefont {Galluzzi}}, \bibinfo {author} {\bibfnamefont {L.}~\bibnamefont {Gozzelino}}, \bibinfo {author} {\bibfnamefont {D.}~\bibnamefont {Torsello}}, \bibinfo {author} {\bibfnamefont {S.}~\bibnamefont {Milošević-Govedarović}}, \bibinfo {author} {\bibfnamefont {J.}~\bibnamefont {Grbović-Novaković}}, \bibinfo {author} {\bibfnamefont {O.~V.}\ \bibnamefont {Dobrovolskiy}}, \bibinfo {author} {\bibfnamefont {W.}~\bibnamefont {Lang}}, \bibinfo {author} {\bibfnamefont {G.}~\bibnamefont {Grimaldi}}, \bibinfo {author} {\bibfnamefont {P.}~\bibnamefont {Crisan}, \bibfnamefont {A.~andd~Badica}}, \bibinfo {author} {\bibfnamefont {A.~M.}\ \bibnamefont {Ionescu}}, \bibinfo {author} {\bibfnamefont {P.}~\bibnamefont {Cayado}}, \bibinfo {author} {\bibfnamefont
  {R.}~\bibnamefont {Willa}}, \bibinfo {author} {\bibfnamefont {B.}~\bibnamefont {Barbiellini}}, \bibinfo {author} {\bibfnamefont {S.}~\bibnamefont {Eley}}, \ and\ \bibinfo {author} {\bibfnamefont {A.}~\bibnamefont {Badía-Majós}},\ }\href {\doibase 10.1016/j.pmatsci.2025.101492} {\bibfield  {journal} {\bibinfo  {journal} {Progress in Materials Science}\ }\textbf {\bibinfo {volume} {155}},\ \bibinfo {pages} {101492} (\bibinfo {year} {2026})}\BibitemShut {NoStop}%
\bibitem [{\citenamefont {Nücker}\ \emph {et~al.}(1988)\citenamefont {Nücker}, \citenamefont {Fink}, \citenamefont {Fuggle}, \citenamefont {Durham},\ and\ \citenamefont {Temmerman}}]{YBCO_hole_doping_O_deficiency}%
  \BibitemOpen
  \bibfield  {author} {\bibinfo {author} {\bibfnamefont {N.}~\bibnamefont {Nücker}}, \bibinfo {author} {\bibfnamefont {J.}~\bibnamefont {Fink}}, \bibinfo {author} {\bibfnamefont {J.~C.}\ \bibnamefont {Fuggle}}, \bibinfo {author} {\bibfnamefont {P.~J.}\ \bibnamefont {Durham}}, \ and\ \bibinfo {author} {\bibfnamefont {W.~M.}\ \bibnamefont {Temmerman}},\ }\href {\doibase 10.1103/PhysRevB.37.5158} {\bibfield  {journal} {\bibinfo  {journal} {Physical Review B}\ }\textbf {\bibinfo {volume} {37}},\ \bibinfo {pages} {5158} (\bibinfo {year} {1988})}\BibitemShut {NoStop}%
\bibitem [{\citenamefont {Arpaia}\ \emph {et~al.}(2019)\citenamefont {Arpaia}, \citenamefont {Caprara}, \citenamefont {Fumagalli}, \citenamefont {De~Vecchi}, \citenamefont {Peng}, \citenamefont {Andersson}, \citenamefont {Betto}, \citenamefont {De~Luca}, \citenamefont {Brookes}, \citenamefont {Lombardi}, \citenamefont {Salluzzo}, \citenamefont {Braicovich}, \citenamefont {Di~Castro}, \citenamefont {Grilli},\ and\ \citenamefont {Ghiringhelli}}]{YBCO_phase_diagram}%
  \BibitemOpen
  \bibfield  {author} {\bibinfo {author} {\bibfnamefont {R.}~\bibnamefont {Arpaia}}, \bibinfo {author} {\bibfnamefont {S.}~\bibnamefont {Caprara}}, \bibinfo {author} {\bibfnamefont {R.}~\bibnamefont {Fumagalli}}, \bibinfo {author} {\bibfnamefont {G.}~\bibnamefont {De~Vecchi}}, \bibinfo {author} {\bibfnamefont {Y.~Y.}\ \bibnamefont {Peng}}, \bibinfo {author} {\bibfnamefont {E.}~\bibnamefont {Andersson}}, \bibinfo {author} {\bibfnamefont {D.}~\bibnamefont {Betto}}, \bibinfo {author} {\bibfnamefont {G.~M.}\ \bibnamefont {De~Luca}}, \bibinfo {author} {\bibfnamefont {N.~B.}\ \bibnamefont {Brookes}}, \bibinfo {author} {\bibfnamefont {F.}~\bibnamefont {Lombardi}}, \bibinfo {author} {\bibfnamefont {M.}~\bibnamefont {Salluzzo}}, \bibinfo {author} {\bibfnamefont {L.}~\bibnamefont {Braicovich}}, \bibinfo {author} {\bibfnamefont {C.}~\bibnamefont {Di~Castro}}, \bibinfo {author} {\bibfnamefont {M.}~\bibnamefont {Grilli}}, \ and\ \bibinfo {author} {\bibfnamefont {G.}~\bibnamefont {Ghiringhelli}},\ }\href {\doibase
  10.1126/science.aav1315} {\bibfield  {journal} {\bibinfo  {journal} {Science}\ }\textbf {\bibinfo {volume} {365}},\ \bibinfo {pages} {906} (\bibinfo {year} {2019})}\BibitemShut {NoStop}%
\bibitem [{\citenamefont {Legris}\ \emph {et~al.}(1993)\citenamefont {Legris}, \citenamefont {Rullier-Albenque}, \citenamefont {Radeva},\ and\ \citenamefont {Lejay}}]{Legris_annealing_1993}%
  \BibitemOpen
  \bibfield  {author} {\bibinfo {author} {\bibfnamefont {A.}~\bibnamefont {Legris}}, \bibinfo {author} {\bibfnamefont {F.}~\bibnamefont {Rullier-Albenque}}, \bibinfo {author} {\bibfnamefont {E.}~\bibnamefont {Radeva}}, \ and\ \bibinfo {author} {\bibfnamefont {P.}~\bibnamefont {Lejay}},\ }\href {\doibase 10.1051/jp1:1993203} {\bibfield  {journal} {\bibinfo  {journal} {Journal de Physique I}\ }\textbf {\bibinfo {volume} {3}},\ \bibinfo {pages} {1605} (\bibinfo {year} {1993})}\BibitemShut {NoStop}%
\bibitem [{\citenamefont {Unterrainer}\ \emph {et~al.}(2022)\citenamefont {Unterrainer}, \citenamefont {Fischer}, \citenamefont {Lorenz},\ and\ \citenamefont {Eisterer}}]{Unterrainer_annealing}%
  \BibitemOpen
  \bibfield  {author} {\bibinfo {author} {\bibfnamefont {R.}~\bibnamefont {Unterrainer}}, \bibinfo {author} {\bibfnamefont {D.~X.}\ \bibnamefont {Fischer}}, \bibinfo {author} {\bibfnamefont {A.}~\bibnamefont {Lorenz}}, \ and\ \bibinfo {author} {\bibfnamefont {M.}~\bibnamefont {Eisterer}},\ }\href {\doibase 10.1088/1361-6668/ac4636} {\bibfield  {journal} {\bibinfo  {journal} {Superconductor Science and Technology}\ }\textbf {\bibinfo {volume} {35}},\ \bibinfo {pages} {04LT01} (\bibinfo {year} {2022})}\BibitemShut {NoStop}%
\bibitem [{\citenamefont {Unterrainer}\ \emph {et~al.}(2024)\citenamefont {Unterrainer}, \citenamefont {Gambino}, \citenamefont {Semper}, \citenamefont {Bodenseher}, \citenamefont {Torsello}, \citenamefont {Laviano}, \citenamefont {Fischer},\ and\ \citenamefont {Eisterer}}]{GdBCO}%
  \BibitemOpen
  \bibfield  {author} {\bibinfo {author} {\bibfnamefont {R.}~\bibnamefont {Unterrainer}}, \bibinfo {author} {\bibfnamefont {D.}~\bibnamefont {Gambino}}, \bibinfo {author} {\bibfnamefont {F.}~\bibnamefont {Semper}}, \bibinfo {author} {\bibfnamefont {A.}~\bibnamefont {Bodenseher}}, \bibinfo {author} {\bibfnamefont {D.}~\bibnamefont {Torsello}}, \bibinfo {author} {\bibfnamefont {F.}~\bibnamefont {Laviano}}, \bibinfo {author} {\bibfnamefont {D.~X.}\ \bibnamefont {Fischer}}, \ and\ \bibinfo {author} {\bibfnamefont {M.}~\bibnamefont {Eisterer}},\ }\href {\doibase 10.1088/1361-6668/ad70db} {\bibfield  {journal} {\bibinfo  {journal} {Superconductor Science and Technology}\ }\textbf {\bibinfo {volume} {37}},\ \bibinfo {pages} {105008} (\bibinfo {year} {2024})}\BibitemShut {NoStop}%
\bibitem [{\citenamefont {Fischer}\ \emph {et~al.}(2025)\citenamefont {Fischer}, \citenamefont {Devitre}, \citenamefont {Woller}, \citenamefont {Fisher}, \citenamefont {Whyte},\ and\ \citenamefont {Hartwig}}]{Fischer_annealing_2025}%
  \BibitemOpen
  \bibfield  {author} {\bibinfo {author} {\bibfnamefont {D.~X.}\ \bibnamefont {Fischer}}, \bibinfo {author} {\bibfnamefont {A.~R.}\ \bibnamefont {Devitre}}, \bibinfo {author} {\bibfnamefont {K.~B.}\ \bibnamefont {Woller}}, \bibinfo {author} {\bibfnamefont {Z.~L.}\ \bibnamefont {Fisher}}, \bibinfo {author} {\bibfnamefont {D.~G.}\ \bibnamefont {Whyte}}, \ and\ \bibinfo {author} {\bibfnamefont {Z.~S.}\ \bibnamefont {Hartwig}},\ }\href {\doibase 10.1088/1361-6668/ada112} {\bibfield  {journal} {\bibinfo  {journal} {Superconductor Science and Technology}\ }\textbf {\bibinfo {volume} {38}},\ \bibinfo {pages} {055019} (\bibinfo {year} {2025})}\BibitemShut {NoStop}%
\bibitem [{\citenamefont {Mundet}\ \emph {et~al.}(2020)\citenamefont {Mundet}, \citenamefont {Hartman}, \citenamefont {Guzman}, \citenamefont {Idrobo}, \citenamefont {Obradors}, \citenamefont {Puig}, \citenamefont {Mishra},\ and\ \citenamefont {Gázquez}}]{Mundet_2020}%
  \BibitemOpen
  \bibfield  {author} {\bibinfo {author} {\bibfnamefont {B.}~\bibnamefont {Mundet}}, \bibinfo {author} {\bibfnamefont {S.~T.}\ \bibnamefont {Hartman}}, \bibinfo {author} {\bibfnamefont {R.}~\bibnamefont {Guzman}}, \bibinfo {author} {\bibfnamefont {J.~C.}\ \bibnamefont {Idrobo}}, \bibinfo {author} {\bibfnamefont {X.}~\bibnamefont {Obradors}}, \bibinfo {author} {\bibfnamefont {T.}~\bibnamefont {Puig}}, \bibinfo {author} {\bibfnamefont {R.}~\bibnamefont {Mishra}}, \ and\ \bibinfo {author} {\bibfnamefont {J.}~\bibnamefont {Gázquez}},\ }\href {\doibase 10.1039/D0NR00666A} {\bibfield  {journal} {\bibinfo  {journal} {Nanoscale}\ }\textbf {\bibinfo {volume} {12}},\ \bibinfo {pages} {5922} (\bibinfo {year} {2020})}\BibitemShut {NoStop}%
\bibitem [{\citenamefont {Torsello}\ \emph {et~al.}(2025{\natexlab{a}})\citenamefont {Torsello}, \citenamefont {Ledda}, \citenamefont {Sparacio}, \citenamefont {Eugenio}, \citenamefont {Giacomo}, \citenamefont {Gallo}, \citenamefont {Hartwig}, \citenamefont {Trotta},\ and\ \citenamefont {Laviano}}]{Torsello_ARC_magnets}%
  \BibitemOpen
  \bibfield  {author} {\bibinfo {author} {\bibfnamefont {D.}~\bibnamefont {Torsello}}, \bibinfo {author} {\bibfnamefont {F.}~\bibnamefont {Ledda}}, \bibinfo {author} {\bibfnamefont {S.}~\bibnamefont {Sparacio}}, \bibinfo {author} {\bibfnamefont {N.~D.}\ \bibnamefont {Eugenio}}, \bibinfo {author} {\bibfnamefont {M.~D.}\ \bibnamefont {Giacomo}}, \bibinfo {author} {\bibfnamefont {E.}~\bibnamefont {Gallo}}, \bibinfo {author} {\bibfnamefont {Z.}~\bibnamefont {Hartwig}}, \bibinfo {author} {\bibfnamefont {A.}~\bibnamefont {Trotta}}, \ and\ \bibinfo {author} {\bibfnamefont {F.}~\bibnamefont {Laviano}},\ }\href {\doibase 10.1109/TASC.2024.3516732} {\bibfield  {journal} {\bibinfo  {journal} {IEEE Transactions on Applied Superconductivity}\ }\textbf {\bibinfo {volume} {35}},\ \bibinfo {pages} {1} (\bibinfo {year} {2025}{\natexlab{a}})}\BibitemShut {NoStop}%
\bibitem [{\citenamefont {Torsello}\ \emph {et~al.}(2023)\citenamefont {Torsello}, \citenamefont {Gambino}, \citenamefont {Gozzelino}, \citenamefont {Trotta},\ and\ \citenamefont {Laviano}}]{Torsello_YBCO_radiation_damage}%
  \BibitemOpen
  \bibfield  {author} {\bibinfo {author} {\bibfnamefont {D.}~\bibnamefont {Torsello}}, \bibinfo {author} {\bibfnamefont {D.}~\bibnamefont {Gambino}}, \bibinfo {author} {\bibfnamefont {L.}~\bibnamefont {Gozzelino}}, \bibinfo {author} {\bibfnamefont {A.}~\bibnamefont {Trotta}}, \ and\ \bibinfo {author} {\bibfnamefont {F.}~\bibnamefont {Laviano}},\ }\href {\doibase 10.1088/1361-6668/aca369} {\bibfield  {journal} {\bibinfo  {journal} {Superconductor Science and Technology}\ }\textbf {\bibinfo {volume} {36}},\ \bibinfo {pages} {014003} (\bibinfo {year} {2023})}\BibitemShut {NoStop}%
\bibitem [{\citenamefont {Torsello}\ \emph {et~al.}(2025{\natexlab{b}})\citenamefont {Torsello}, \citenamefont {Celentano}, \citenamefont {Civale}, \citenamefont {Corato}, \citenamefont {Eisterer}, \citenamefont {Gambino}, \citenamefont {Murphy}, \citenamefont {Speller},\ and\ \citenamefont {Laviano}}]{HTS_fusion_roadmap_2025}%
  \BibitemOpen
  \bibfield  {author} {\bibinfo {author} {\bibfnamefont {D.}~\bibnamefont {Torsello}}, \bibinfo {author} {\bibfnamefont {G.}~\bibnamefont {Celentano}}, \bibinfo {author} {\bibfnamefont {L.}~\bibnamefont {Civale}}, \bibinfo {author} {\bibfnamefont {V.}~\bibnamefont {Corato}}, \bibinfo {author} {\bibfnamefont {M.}~\bibnamefont {Eisterer}}, \bibinfo {author} {\bibfnamefont {D.}~\bibnamefont {Gambino}}, \bibinfo {author} {\bibfnamefont {S.}~\bibnamefont {Murphy}}, \bibinfo {author} {\bibfnamefont {S.}~\bibnamefont {Speller}}, \ and\ \bibinfo {author} {\bibfnamefont {F.}~\bibnamefont {Laviano}},\ }\href {\doibase 10.1088/1361-6668/adce40} {\bibfield  {journal} {\bibinfo  {journal} {Superconductor Science and Technology}\ }\textbf {\bibinfo {volume} {38}},\ \bibinfo {pages} {053501} (\bibinfo {year} {2025}{\natexlab{b}})}\BibitemShut {NoStop}%
\bibitem [{\citenamefont {Jorgensen}\ \emph {et~al.}(1987)\citenamefont {Jorgensen}, \citenamefont {Beno}, \citenamefont {Hinks}, \citenamefont {Soderholm}, \citenamefont {Volin}, \citenamefont {Hitterman}, \citenamefont {Grace}, \citenamefont {Schuller}, \citenamefont {Segre}, \citenamefont {Zhang},\ and\ \citenamefont {Kleefisch}}]{expTransition}%
  \BibitemOpen
  \bibfield  {author} {\bibinfo {author} {\bibfnamefont {J.~D.}\ \bibnamefont {Jorgensen}}, \bibinfo {author} {\bibfnamefont {M.~A.}\ \bibnamefont {Beno}}, \bibinfo {author} {\bibfnamefont {D.~G.}\ \bibnamefont {Hinks}}, \bibinfo {author} {\bibfnamefont {L.}~\bibnamefont {Soderholm}}, \bibinfo {author} {\bibfnamefont {K.~J.}\ \bibnamefont {Volin}}, \bibinfo {author} {\bibfnamefont {R.~L.}\ \bibnamefont {Hitterman}}, \bibinfo {author} {\bibfnamefont {J.~D.}\ \bibnamefont {Grace}}, \bibinfo {author} {\bibfnamefont {I.~K.}\ \bibnamefont {Schuller}}, \bibinfo {author} {\bibfnamefont {C.~U.}\ \bibnamefont {Segre}}, \bibinfo {author} {\bibfnamefont {K.}~\bibnamefont {Zhang}}, \ and\ \bibinfo {author} {\bibfnamefont {M.~S.}\ \bibnamefont {Kleefisch}},\ }\href {\doibase 10.1103/PhysRevB.36.3608} {\bibfield  {journal} {\bibinfo  {journal} {Phys. Rev. B}\ }\textbf {\bibinfo {volume} {36}},\ \bibinfo {pages} {3608} (\bibinfo {year} {1987})}\BibitemShut {NoStop}%
\bibitem [{\citenamefont {Schuller}\ \emph {et~al.}(1987)\citenamefont {Schuller}, \citenamefont {Hinks}, \citenamefont {Beno}, \citenamefont {Capone}, \citenamefont {Soderholm}, \citenamefont {Locquet}, \citenamefont {Bruynseraede}, \citenamefont {Segre},\ and\ \citenamefont {Zhang}}]{ExpTransitionAndSC}%
  \BibitemOpen
  \bibfield  {author} {\bibinfo {author} {\bibfnamefont {I.~K.}\ \bibnamefont {Schuller}}, \bibinfo {author} {\bibfnamefont {D.}~\bibnamefont {Hinks}}, \bibinfo {author} {\bibfnamefont {M.}~\bibnamefont {Beno}}, \bibinfo {author} {\bibfnamefont {D.}~\bibnamefont {Capone}}, \bibinfo {author} {\bibfnamefont {L.}~\bibnamefont {Soderholm}}, \bibinfo {author} {\bibfnamefont {J.-P.}\ \bibnamefont {Locquet}}, \bibinfo {author} {\bibfnamefont {Y.}~\bibnamefont {Bruynseraede}}, \bibinfo {author} {\bibfnamefont {C.}~\bibnamefont {Segre}}, \ and\ \bibinfo {author} {\bibfnamefont {K.}~\bibnamefont {Zhang}},\ }\href {\doibase https://doi.org/10.1016/0038-1098(87)91134-3} {\bibfield  {journal} {\bibinfo  {journal} {Solid State Communications}\ }\textbf {\bibinfo {volume} {63}},\ \bibinfo {pages} {385} (\bibinfo {year} {1987})}\BibitemShut {NoStop}%
\bibitem [{\citenamefont {Bonetti}\ \emph {et~al.}(1994)\citenamefont {Bonetti}, \citenamefont {Campari}, \citenamefont {Mattioli},\ and\ \citenamefont {Zingaro}}]{expTransition_vs_O_content}%
  \BibitemOpen
  \bibfield  {author} {\bibinfo {author} {\bibfnamefont {E.}~\bibnamefont {Bonetti}}, \bibinfo {author} {\bibfnamefont {E.~G.}\ \bibnamefont {Campari}}, \bibinfo {author} {\bibfnamefont {P.}~\bibnamefont {Mattioli}}, \ and\ \bibinfo {author} {\bibfnamefont {A.}~\bibnamefont {Zingaro}},\ }\href {\doibase 10.1016/0925-8388(94)90510-X} {\bibfield  {journal} {\bibinfo  {journal} {Journal of Alloys and Compounds}\ }\bibinfo {series} {10th {International} {Conference} on {Internal} {Friction} and {Ultrasonic} {Attenuation} in {Solids}},\ \textbf {\bibinfo {volume} {211-212}},\ \bibinfo {pages} {314} (\bibinfo {year} {1994})}\BibitemShut {NoStop}%
\bibitem [{\citenamefont {Chaplot}(1989)}]{Chaplot_potential_transition}%
  \BibitemOpen
  \bibfield  {author} {\bibinfo {author} {\bibfnamefont {S.~L.}\ \bibnamefont {Chaplot}},\ }\href {\doibase 10.1080/01411598908242380} {\bibfield  {journal} {\bibinfo  {journal} {Phase Transitions}\ }\textbf {\bibinfo {volume} {19}},\ \bibinfo {pages} {49} (\bibinfo {year} {1989})}\BibitemShut {NoStop}%
\bibitem [{\citenamefont {Chaplot}(1988)}]{Chaplot_potential_phonons}%
  \BibitemOpen
  \bibfield  {author} {\bibinfo {author} {\bibfnamefont {S.~L.}\ \bibnamefont {Chaplot}},\ }\href {\doibase 10.1103/PhysRevB.37.7435} {\bibfield  {journal} {\bibinfo  {journal} {Phys. Rev. B}\ }\textbf {\bibinfo {volume} {37}},\ \bibinfo {pages} {7435} (\bibinfo {year} {1988})}\BibitemShut {NoStop}%
\bibitem [{\citenamefont {Chaplot}(1990)}]{Chaplot_potential_transition_and_phonons}%
  \BibitemOpen
  \bibfield  {author} {\bibinfo {author} {\bibfnamefont {S.~L.}\ \bibnamefont {Chaplot}},\ }\href {\doibase 10.1103/PhysRevB.42.2149} {\bibfield  {journal} {\bibinfo  {journal} {Phys. Rev. B}\ }\textbf {\bibinfo {volume} {42}},\ \bibinfo {pages} {2149} (\bibinfo {year} {1990})}\BibitemShut {NoStop}%
\bibitem [{\citenamefont {Gray}\ \emph {et~al.}(2022)\citenamefont {Gray}, \citenamefont {Rushton},\ and\ \citenamefont {Murphy}}]{GrayPotential}%
  \BibitemOpen
  \bibfield  {author} {\bibinfo {author} {\bibfnamefont {R.~L.}\ \bibnamefont {Gray}}, \bibinfo {author} {\bibfnamefont {M.~J.~D.}\ \bibnamefont {Rushton}}, \ and\ \bibinfo {author} {\bibfnamefont {S.~T.}\ \bibnamefont {Murphy}},\ }\href {\doibase 10.1088/1361-6668/ac47dc} {\bibfield  {journal} {\bibinfo  {journal} {Superconductor Science and Technology}\ }\textbf {\bibinfo {volume} {35}},\ \bibinfo {pages} {035010} (\bibinfo {year} {2022})}\BibitemShut {NoStop}%
\bibitem [{\citenamefont {Dickson}\ \emph {et~al.}(2025)\citenamefont {Dickson}, \citenamefont {Gilbert}, \citenamefont {Nguyen-Manh},\ and\ \citenamefont {Murphy}}]{Dickson_TDE}%
  \BibitemOpen
  \bibfield  {author} {\bibinfo {author} {\bibfnamefont {A.}~\bibnamefont {Dickson}}, \bibinfo {author} {\bibfnamefont {M.~R.}\ \bibnamefont {Gilbert}}, \bibinfo {author} {\bibfnamefont {D.}~\bibnamefont {Nguyen-Manh}}, \ and\ \bibinfo {author} {\bibfnamefont {S.~T.}\ \bibnamefont {Murphy}},\ }\href {\doibase 10.48550/arXiv.2506.17976} {{\bibinfo {title} {\tt arXiv:2506.17976 [cond-mat.supr-con]},}\ } (\bibinfo {year} {2025})\BibitemShut {NoStop}%
\bibitem [{\citenamefont {Behler}(2016)}]{ML_general_I}%
  \BibitemOpen
  \bibfield  {author} {\bibinfo {author} {\bibfnamefont {J.}~\bibnamefont {Behler}},\ }\href {\doibase 10.1063/1.4966192} {\bibfield  {journal} {\bibinfo  {journal} {The Journal of Chemical Physics}\ }\textbf {\bibinfo {volume} {145}},\ \bibinfo {pages} {170901} (\bibinfo {year} {2016})}\BibitemShut {NoStop}%
\bibitem [{\citenamefont {Friederich}\ \emph {et~al.}(2021)\citenamefont {Friederich}, \citenamefont {Häse}, \citenamefont {Proppe},\ and\ \citenamefont {Aspuru-Guzik}}]{ML_general_II}%
  \BibitemOpen
  \bibfield  {author} {\bibinfo {author} {\bibfnamefont {P.}~\bibnamefont {Friederich}}, \bibinfo {author} {\bibfnamefont {F.}~\bibnamefont {Häse}}, \bibinfo {author} {\bibfnamefont {J.}~\bibnamefont {Proppe}}, \ and\ \bibinfo {author} {\bibfnamefont {A.}~\bibnamefont {Aspuru-Guzik}},\ }\href {\doibase 10.1038/s41563-020-0777-6} {\bibfield  {journal} {\bibinfo  {journal} {Nature Materials}\ }\textbf {\bibinfo {volume} {20}},\ \bibinfo {pages} {750} (\bibinfo {year} {2021})}\BibitemShut {NoStop}%
\bibitem [{\citenamefont {Mishin}(2021)}]{ML_general_III}%
  \BibitemOpen
  \bibfield  {author} {\bibinfo {author} {\bibfnamefont {Y.}~\bibnamefont {Mishin}},\ }\href {\doibase 10.1016/j.actamat.2021.116980} {\bibfield  {journal} {\bibinfo  {journal} {Acta Materialia}\ }\textbf {\bibinfo {volume} {214}},\ \bibinfo {pages} {116980} (\bibinfo {year} {2021})}\BibitemShut {NoStop}%
\bibitem [{\citenamefont {Deringer}\ \emph {et~al.}(2019)\citenamefont {Deringer}, \citenamefont {Caro},\ and\ \citenamefont {Csányi}}]{ML_general_IV}%
  \BibitemOpen
  \bibfield  {author} {\bibinfo {author} {\bibfnamefont {V.~L.}\ \bibnamefont {Deringer}}, \bibinfo {author} {\bibfnamefont {M.~A.}\ \bibnamefont {Caro}}, \ and\ \bibinfo {author} {\bibfnamefont {G.}~\bibnamefont {Csányi}},\ }\href {\doibase 10.1002/adma.201902765} {\bibfield  {journal} {\bibinfo  {journal} {Advanced Materials}\ }\textbf {\bibinfo {volume} {31}},\ \bibinfo {pages} {1902765} (\bibinfo {year} {2019})}\BibitemShut {NoStop}%
\bibitem [{\citenamefont {Bochkarev}\ \emph {et~al.}(2022)\citenamefont {Bochkarev}, \citenamefont {Lysogorskiy}, \citenamefont {Menon}, \citenamefont {Qamar}, \citenamefont {Mrovec},\ and\ \citenamefont {Drautz}}]{Pacemaker_PRM}%
  \BibitemOpen
  \bibfield  {author} {\bibinfo {author} {\bibfnamefont {A.}~\bibnamefont {Bochkarev}}, \bibinfo {author} {\bibfnamefont {Y.}~\bibnamefont {Lysogorskiy}}, \bibinfo {author} {\bibfnamefont {S.}~\bibnamefont {Menon}}, \bibinfo {author} {\bibfnamefont {M.}~\bibnamefont {Qamar}}, \bibinfo {author} {\bibfnamefont {M.}~\bibnamefont {Mrovec}}, \ and\ \bibinfo {author} {\bibfnamefont {R.}~\bibnamefont {Drautz}},\ }\href {\doibase 10.1103/PhysRevMaterials.6.013804} {\bibfield  {journal} {\bibinfo  {journal} {Phys. Rev. Mater.}\ }\textbf {\bibinfo {volume} {6}},\ \bibinfo {pages} {013804} (\bibinfo {year} {2022})}\BibitemShut {NoStop}%
\bibitem [{\citenamefont {Loutati}\ \emph {et~al.}(2021)\citenamefont {Loutati}, \citenamefont {Sohn},\ and\ \citenamefont {Tietz}}]{ACE_NaZrSiPO}%
  \BibitemOpen
  \bibfield  {author} {\bibinfo {author} {\bibfnamefont {A.}~\bibnamefont {Loutati}}, \bibinfo {author} {\bibfnamefont {Y.~J.}\ \bibnamefont {Sohn}}, \ and\ \bibinfo {author} {\bibfnamefont {F.}~\bibnamefont {Tietz}},\ }\href {\doibase 10.1002/cphc.202100032} {\bibfield  {journal} {\bibinfo  {journal} {ChemPhysChem}\ }\textbf {\bibinfo {volume} {22}},\ \bibinfo {pages} {995} (\bibinfo {year} {2021})}\BibitemShut {NoStop}%
\bibitem [{\citenamefont {Attarian}\ \emph {et~al.}(2024)\citenamefont {Attarian}, \citenamefont {Morgan},\ and\ \citenamefont {Szlufarska}}]{ACE_NiCrFLiBe}%
  \BibitemOpen
  \bibfield  {author} {\bibinfo {author} {\bibfnamefont {S.}~\bibnamefont {Attarian}}, \bibinfo {author} {\bibfnamefont {D.}~\bibnamefont {Morgan}}, \ and\ \bibinfo {author} {\bibfnamefont {I.}~\bibnamefont {Szlufarska}},\ }\href {\doibase 10.1016/j.molliq.2024.124521} {\bibfield  {journal} {\bibinfo  {journal} {Journal of Molecular Liquids}\ }\textbf {\bibinfo {volume} {400}},\ \bibinfo {pages} {124521} (\bibinfo {year} {2024})}\BibitemShut {NoStop}%
\bibitem [{\citenamefont {Liu}\ \emph {et~al.}(2023{\natexlab{a}})\citenamefont {Liu}, \citenamefont {Byggm\"astar}, \citenamefont {Fan}, \citenamefont {Qian},\ and\ \citenamefont {Su}}]{ML_W_defs}%
  \BibitemOpen
  \bibfield  {author} {\bibinfo {author} {\bibfnamefont {J.}~\bibnamefont {Liu}}, \bibinfo {author} {\bibfnamefont {J.}~\bibnamefont {Byggm\"astar}}, \bibinfo {author} {\bibfnamefont {Z.}~\bibnamefont {Fan}}, \bibinfo {author} {\bibfnamefont {P.}~\bibnamefont {Qian}}, \ and\ \bibinfo {author} {\bibfnamefont {Y.}~\bibnamefont {Su}},\ }\href {\doibase 10.1103/PhysRevB.108.054312} {\bibfield  {journal} {\bibinfo  {journal} {Phys. Rev. B}\ }\textbf {\bibinfo {volume} {108}},\ \bibinfo {pages} {054312} (\bibinfo {year} {2023}{\natexlab{a}})}\BibitemShut {NoStop}%
\bibitem [{\citenamefont {Leimeroth}\ \emph {et~al.}(2024)\citenamefont {Leimeroth}, \citenamefont {Rohrer},\ and\ \citenamefont {Albe}}]{ACE_CuZr_defs}%
  \BibitemOpen
  \bibfield  {author} {\bibinfo {author} {\bibfnamefont {N.}~\bibnamefont {Leimeroth}}, \bibinfo {author} {\bibfnamefont {J.}~\bibnamefont {Rohrer}}, \ and\ \bibinfo {author} {\bibfnamefont {K.}~\bibnamefont {Albe}},\ }\href {\doibase 10.1103/PhysRevMaterials.8.043602} {\bibfield  {journal} {\bibinfo  {journal} {Phys. Rev. Mater.}\ }\textbf {\bibinfo {volume} {8}},\ \bibinfo {pages} {043602} (\bibinfo {year} {2024})}\BibitemShut {NoStop}%
\bibitem [{\citenamefont {Liu}\ \emph {et~al.}(2023{\natexlab{b}})\citenamefont {Liu}, \citenamefont {He},\ and\ \citenamefont {Mo}}]{ML_discrepancies_defs}%
  \BibitemOpen
  \bibfield  {author} {\bibinfo {author} {\bibfnamefont {Y.}~\bibnamefont {Liu}}, \bibinfo {author} {\bibfnamefont {X.}~\bibnamefont {He}}, \ and\ \bibinfo {author} {\bibfnamefont {Y.}~\bibnamefont {Mo}},\ }\href {\doibase 10.1038/s41524-023-01123-3} {\bibfield  {journal} {\bibinfo  {journal} {npj Computational Materials}\ }\textbf {\bibinfo {volume} {9}},\ \bibinfo {pages} {174} (\bibinfo {year} {2023}{\natexlab{b}})}\BibitemShut {NoStop}%
\bibitem [{\citenamefont {Fellman}\ \emph {et~al.}(2025)\citenamefont {Fellman}, \citenamefont {Byggmästar}, \citenamefont {Granberg}, \citenamefont {Nordlund},\ and\ \citenamefont {Djurabekova}}]{ML_Cu_Al_Ni_defs}%
  \BibitemOpen
  \bibfield  {author} {\bibinfo {author} {\bibfnamefont {A.}~\bibnamefont {Fellman}}, \bibinfo {author} {\bibfnamefont {J.}~\bibnamefont {Byggmästar}}, \bibinfo {author} {\bibfnamefont {F.}~\bibnamefont {Granberg}}, \bibinfo {author} {\bibfnamefont {K.}~\bibnamefont {Nordlund}}, \ and\ \bibinfo {author} {\bibfnamefont {F.}~\bibnamefont {Djurabekova}},\ }\href {\doibase 10.1103/PhysRevMaterials.9.053807} {\bibfield  {journal} {\bibinfo  {journal} {Physical Review Materials}\ }\textbf {\bibinfo {volume} {9}},\ \bibinfo {pages} {053807} (\bibinfo {year} {2025})}\BibitemShut {NoStop}%
\bibitem [{\citenamefont {Poul}\ \emph {et~al.}(2025)\citenamefont {Poul}, \citenamefont {Huber},\ and\ \citenamefont {Neugebauer}}]{ML_MPIE_defs}%
  \BibitemOpen
  \bibfield  {author} {\bibinfo {author} {\bibfnamefont {M.}~\bibnamefont {Poul}}, \bibinfo {author} {\bibfnamefont {L.}~\bibnamefont {Huber}}, \ and\ \bibinfo {author} {\bibfnamefont {J.}~\bibnamefont {Neugebauer}},\ }\href {\doibase 10.1038/s41524-025-01669-4} {\bibfield  {journal} {\bibinfo  {journal} {npj Computational Materials}\ }\textbf {\bibinfo {volume} {11}},\ \bibinfo {pages} {174} (\bibinfo {year} {2025})}\BibitemShut {NoStop}%
\bibitem [{\citenamefont {Drautz}(2019)}]{ACE_Drautz_I}%
  \BibitemOpen
  \bibfield  {author} {\bibinfo {author} {\bibfnamefont {R.}~\bibnamefont {Drautz}},\ }\href {\doibase 10.1103/PhysRevB.99.014104} {\bibfield  {journal} {\bibinfo  {journal} {Phys. Rev. B}\ }\textbf {\bibinfo {volume} {99}},\ \bibinfo {pages} {014104} (\bibinfo {year} {2019})}\BibitemShut {NoStop}%
\bibitem [{\citenamefont {Drautz}(2020)}]{ACE_Drautz_II}%
  \BibitemOpen
  \bibfield  {author} {\bibinfo {author} {\bibfnamefont {R.}~\bibnamefont {Drautz}},\ }\href {\doibase 10.1103/PhysRevB.102.024104} {\bibfield  {journal} {\bibinfo  {journal} {Physical Review B}\ }\textbf {\bibinfo {volume} {102}},\ \bibinfo {pages} {024104} (\bibinfo {year} {2020})}\BibitemShut {NoStop}%
\bibitem [{\citenamefont {Stukowski}(2009)}]{ovito}%
  \BibitemOpen
  \bibfield  {author} {\bibinfo {author} {\bibfnamefont {A.}~\bibnamefont {Stukowski}},\ }\href {\doibase 10.1088/0965-0393/18/1/015012} {\bibfield  {journal} {\bibinfo  {journal} {Modelling and Simulation in Materials Science and Engineering}\ }\textbf {\bibinfo {volume} {18}},\ \bibinfo {pages} {015012} (\bibinfo {year} {2009})}\BibitemShut {NoStop}%
\bibitem [{\citenamefont {Ziegler}\ and\ \citenamefont {Biersack}(1985)}]{ZBL}%
  \BibitemOpen
  \bibfield  {author} {\bibinfo {author} {\bibfnamefont {J.~F.}\ \bibnamefont {Ziegler}}\ and\ \bibinfo {author} {\bibfnamefont {J.~P.}\ \bibnamefont {Biersack}},\ }\enquote {\bibinfo {title} {The stopping and range of ions in matter},}\ in\ \href {\doibase 10.1007/978-1-4615-8103-1_3} {\emph {\bibinfo {booktitle} {Treatise on Heavy-Ion Science: Volume 6: Astrophysics, Chemistry, and Condensed Matter}}},\ \bibinfo {editor} {edited by\ \bibinfo {editor} {\bibfnamefont {D.~A.}\ \bibnamefont {Bromley}}}\ (\bibinfo  {publisher} {Springer US},\ \bibinfo {address} {Boston, MA},\ \bibinfo {year} {1985})\ pp.\ \bibinfo {pages} {93--129}\BibitemShut {NoStop}%
\bibitem [{\citenamefont {Hjorth~Larsen}\ \emph {et~al.}(2017)\citenamefont {Hjorth~Larsen}, \citenamefont {J{\o}rgen~Mortensen}, \citenamefont {Blomqvist}, \citenamefont {Castelli}, \citenamefont {Christensen}, \citenamefont {Du{\l}ak}, \citenamefont {Friis}, \citenamefont {Groves}, \citenamefont {Hammer}, \citenamefont {Hargus}, \citenamefont {Hermes}, \citenamefont {Jennings}, \citenamefont {Bjerre~Jensen}, \citenamefont {Kermode}, \citenamefont {Kitchin}, \citenamefont {Leonhard~Kolsbjerg}, \citenamefont {Kubal}, \citenamefont {Kaasbjerg}, \citenamefont {Lysgaard}, \citenamefont {Bergmann~Maronsson}, \citenamefont {Maxson}, \citenamefont {Ollsen}, \citenamefont {Pastewka}, \citenamefont {Peterson}, \citenamefont {Rostgaard}, \citenamefont {Schi{\o}tz}, \citenamefont {Sch{\"u}tt}, \citenamefont {Strange}, \citenamefont {Thygesen}, \citenamefont {Vegge}, \citenamefont {Vilhelmsen}, \citenamefont {Walter}, \citenamefont {Zeng},\ and\ \citenamefont {Jacobsen}}]{ASE}%
  \BibitemOpen
  \bibfield  {author} {\bibinfo {author} {\bibfnamefont {A.}~\bibnamefont {Hjorth~Larsen}}, \bibinfo {author} {\bibfnamefont {J.}~\bibnamefont {J{\o}rgen~Mortensen}}, \bibinfo {author} {\bibfnamefont {J.}~\bibnamefont {Blomqvist}}, \bibinfo {author} {\bibfnamefont {I.~E.}\ \bibnamefont {Castelli}}, \bibinfo {author} {\bibfnamefont {R.}~\bibnamefont {Christensen}}, \bibinfo {author} {\bibfnamefont {M.}~\bibnamefont {Du{\l}ak}}, \bibinfo {author} {\bibfnamefont {J.}~\bibnamefont {Friis}}, \bibinfo {author} {\bibfnamefont {M.~N.}\ \bibnamefont {Groves}}, \bibinfo {author} {\bibfnamefont {B.}~\bibnamefont {Hammer}}, \bibinfo {author} {\bibfnamefont {C.}~\bibnamefont {Hargus}}, \bibinfo {author} {\bibfnamefont {E.~D.}\ \bibnamefont {Hermes}}, \bibinfo {author} {\bibfnamefont {P.~C.}\ \bibnamefont {Jennings}}, \bibinfo {author} {\bibfnamefont {P.}~\bibnamefont {Bjerre~Jensen}}, \bibinfo {author} {\bibfnamefont {J.}~\bibnamefont {Kermode}}, \bibinfo {author} {\bibfnamefont {J.~R.}\ \bibnamefont {Kitchin}}, \bibinfo
  {author} {\bibfnamefont {E.}~\bibnamefont {Leonhard~Kolsbjerg}}, \bibinfo {author} {\bibfnamefont {J.}~\bibnamefont {Kubal}}, \bibinfo {author} {\bibfnamefont {K.}~\bibnamefont {Kaasbjerg}}, \bibinfo {author} {\bibfnamefont {S.}~\bibnamefont {Lysgaard}}, \bibinfo {author} {\bibfnamefont {J.}~\bibnamefont {Bergmann~Maronsson}}, \bibinfo {author} {\bibfnamefont {T.}~\bibnamefont {Maxson}}, \bibinfo {author} {\bibfnamefont {T.}~\bibnamefont {Ollsen}}, \bibinfo {author} {\bibfnamefont {L.}~\bibnamefont {Pastewka}}, \bibinfo {author} {\bibfnamefont {A.}~\bibnamefont {Peterson}}, \bibinfo {author} {\bibfnamefont {C.}~\bibnamefont {Rostgaard}}, \bibinfo {author} {\bibfnamefont {J.}~\bibnamefont {Schi{\o}tz}}, \bibinfo {author} {\bibfnamefont {O.}~\bibnamefont {Sch{\"u}tt}}, \bibinfo {author} {\bibfnamefont {M.}~\bibnamefont {Strange}}, \bibinfo {author} {\bibfnamefont {K.~S.}\ \bibnamefont {Thygesen}}, \bibinfo {author} {\bibfnamefont {T.}~\bibnamefont {Vegge}}, \bibinfo {author} {\bibfnamefont {L.}~\bibnamefont
  {Vilhelmsen}}, \bibinfo {author} {\bibfnamefont {M.}~\bibnamefont {Walter}}, \bibinfo {author} {\bibfnamefont {Z.}~\bibnamefont {Zeng}}, \ and\ \bibinfo {author} {\bibfnamefont {K.~W.}\ \bibnamefont {Jacobsen}},\ }\href {\doibase 10.1088/1361-648X/aa680e} {\bibfield  {journal} {\bibinfo  {journal} {Journal of Physics: Condensed Matter}\ }\textbf {\bibinfo {volume} {29}},\ \bibinfo {pages} {273002} (\bibinfo {year} {2017})}\BibitemShut {NoStop}%
\bibitem [{\citenamefont {Hirel}(2015)}]{Atomsk}%
  \BibitemOpen
  \bibfield  {author} {\bibinfo {author} {\bibfnamefont {P.}~\bibnamefont {Hirel}},\ }\href {\doibase 10.1016/j.cpc.2015.07.012} {\bibfield  {journal} {\bibinfo  {journal} {Computer Physics Communications}\ }\textbf {\bibinfo {volume} {197}},\ \bibinfo {pages} {212} (\bibinfo {year} {2015})}\BibitemShut {NoStop}%
\bibitem [{\citenamefont {Kresse}\ and\ \citenamefont {Hafner}(1993)}]{VASP_I}%
  \BibitemOpen
  \bibfield  {author} {\bibinfo {author} {\bibfnamefont {G.}~\bibnamefont {Kresse}}\ and\ \bibinfo {author} {\bibfnamefont {J.}~\bibnamefont {Hafner}},\ }\href {\doibase 10.1103/PhysRevB.47.558} {\bibfield  {journal} {\bibinfo  {journal} {Phys. Rev. B}\ }\textbf {\bibinfo {volume} {47}},\ \bibinfo {pages} {558} (\bibinfo {year} {1993})}\BibitemShut {NoStop}%
\bibitem [{\citenamefont {Kresse}\ and\ \citenamefont {Furthm\"uller}(1996{\natexlab{a}})}]{VASP_II}%
  \BibitemOpen
  \bibfield  {author} {\bibinfo {author} {\bibfnamefont {G.}~\bibnamefont {Kresse}}\ and\ \bibinfo {author} {\bibfnamefont {J.}~\bibnamefont {Furthm\"uller}},\ }\href {\doibase https://doi.org/10.1016/0927-0256(96)00008-0} {\bibfield  {journal} {\bibinfo  {journal} {Computational Materials Science}\ }\textbf {\bibinfo {volume} {6}},\ \bibinfo {pages} {15 } (\bibinfo {year} {1996}{\natexlab{a}})}\BibitemShut {NoStop}%
\bibitem [{\citenamefont {Kresse}\ and\ \citenamefont {Furthm\"uller}(1996{\natexlab{b}})}]{VASP_III}%
  \BibitemOpen
  \bibfield  {author} {\bibinfo {author} {\bibfnamefont {G.}~\bibnamefont {Kresse}}\ and\ \bibinfo {author} {\bibfnamefont {J.}~\bibnamefont {Furthm\"uller}},\ }\href {\doibase 10.1103/PhysRevB.54.11169} {\bibfield  {journal} {\bibinfo  {journal} {Phys. Rev. B}\ }\textbf {\bibinfo {volume} {54}},\ \bibinfo {pages} {11169} (\bibinfo {year} {1996}{\natexlab{b}})}\BibitemShut {NoStop}%
\bibitem [{\citenamefont {Bl\"ochl}(1994)}]{PAW_Blochl}%
  \BibitemOpen
  \bibfield  {author} {\bibinfo {author} {\bibfnamefont {P.~E.}\ \bibnamefont {Bl\"ochl}},\ }\href {\doibase 10.1103/PhysRevB.50.17953} {\bibfield  {journal} {\bibinfo  {journal} {Phys. Rev. B}\ }\textbf {\bibinfo {volume} {50}},\ \bibinfo {pages} {17953} (\bibinfo {year} {1994})}\BibitemShut {NoStop}%
\bibitem [{\citenamefont {Kresse}\ and\ \citenamefont {Joubert}(1999)}]{PAW_vasp}%
  \BibitemOpen
  \bibfield  {author} {\bibinfo {author} {\bibfnamefont {G.}~\bibnamefont {Kresse}}\ and\ \bibinfo {author} {\bibfnamefont {D.}~\bibnamefont {Joubert}},\ }\href {\doibase 10.1103/PhysRevB.59.1758} {\bibfield  {journal} {\bibinfo  {journal} {Phys. Rev. B}\ }\textbf {\bibinfo {volume} {59}},\ \bibinfo {pages} {1758} (\bibinfo {year} {1999})}\BibitemShut {NoStop}%
\bibitem [{\citenamefont {Perdew}\ \emph {et~al.}(1996)\citenamefont {Perdew}, \citenamefont {Burke},\ and\ \citenamefont {Ernzerhof}}]{PBE}%
  \BibitemOpen
  \bibfield  {author} {\bibinfo {author} {\bibfnamefont {J.~P.}\ \bibnamefont {Perdew}}, \bibinfo {author} {\bibfnamefont {K.}~\bibnamefont {Burke}}, \ and\ \bibinfo {author} {\bibfnamefont {M.}~\bibnamefont {Ernzerhof}},\ }\href {\doibase 10.1103/PhysRevLett.77.3865} {\bibfield  {journal} {\bibinfo  {journal} {Physical Review Letters}\ }\textbf {\bibinfo {volume} {77}},\ \bibinfo {pages} {3865} (\bibinfo {year} {1996})}\BibitemShut {NoStop}%
\bibitem [{\citenamefont {Monkhorst}\ and\ \citenamefont {Pack}(1976)}]{MPscheme}%
  \BibitemOpen
  \bibfield  {author} {\bibinfo {author} {\bibfnamefont {H.~J.}\ \bibnamefont {Monkhorst}}\ and\ \bibinfo {author} {\bibfnamefont {J.~D.}\ \bibnamefont {Pack}},\ }\href {\doibase 10.1103/PhysRevB.13.5188} {\bibfield  {journal} {\bibinfo  {journal} {Phys. Rev. B}\ }\textbf {\bibinfo {volume} {13}},\ \bibinfo {pages} {5188} (\bibinfo {year} {1976})}\BibitemShut {NoStop}%
\bibitem [{\citenamefont {Lysogorskiy}\ \emph {et~al.}(2021)\citenamefont {Lysogorskiy}, \citenamefont {Oord}, \citenamefont {Bochkarev}, \citenamefont {Menon}, \citenamefont {Rinaldi}, \citenamefont {Hammerschmidt}, \citenamefont {Mrovec}, \citenamefont {Thompson}, \citenamefont {Csányi}, \citenamefont {Ortner},\ and\ \citenamefont {Drautz}}]{Pacemaker_NPJ}%
  \BibitemOpen
  \bibfield  {author} {\bibinfo {author} {\bibfnamefont {Y.}~\bibnamefont {Lysogorskiy}}, \bibinfo {author} {\bibfnamefont {C.~v.~d.}\ \bibnamefont {Oord}}, \bibinfo {author} {\bibfnamefont {A.}~\bibnamefont {Bochkarev}}, \bibinfo {author} {\bibfnamefont {S.}~\bibnamefont {Menon}}, \bibinfo {author} {\bibfnamefont {M.}~\bibnamefont {Rinaldi}}, \bibinfo {author} {\bibfnamefont {T.}~\bibnamefont {Hammerschmidt}}, \bibinfo {author} {\bibfnamefont {M.}~\bibnamefont {Mrovec}}, \bibinfo {author} {\bibfnamefont {A.}~\bibnamefont {Thompson}}, \bibinfo {author} {\bibfnamefont {G.}~\bibnamefont {Csányi}}, \bibinfo {author} {\bibfnamefont {C.}~\bibnamefont {Ortner}}, \ and\ \bibinfo {author} {\bibfnamefont {R.}~\bibnamefont {Drautz}},\ }\href {\doibase 10.1038/s41524-021-00559-9} {\bibfield  {journal} {\bibinfo  {journal} {npj Computational Materials}\ }\textbf {\bibinfo {volume} {7}},\ \bibinfo {pages} {1} (\bibinfo {year} {2021})}\BibitemShut {NoStop}%
\bibitem [{\citenamefont {Thompson}\ \emph {et~al.}(2022)\citenamefont {Thompson}, \citenamefont {Aktulga}, \citenamefont {Berger}, \citenamefont {Bolintineanu}, \citenamefont {Brown}, \citenamefont {Crozier}, \citenamefont {in~’t Veld}, \citenamefont {Kohlmeyer}, \citenamefont {Moore}, \citenamefont {Nguyen}, \citenamefont {Shan}, \citenamefont {Stevens}, \citenamefont {Tranchida}, \citenamefont {Trott},\ and\ \citenamefont {Plimpton}}]{lammps}%
  \BibitemOpen
  \bibfield  {author} {\bibinfo {author} {\bibfnamefont {A.~P.}\ \bibnamefont {Thompson}}, \bibinfo {author} {\bibfnamefont {H.~M.}\ \bibnamefont {Aktulga}}, \bibinfo {author} {\bibfnamefont {R.}~\bibnamefont {Berger}}, \bibinfo {author} {\bibfnamefont {D.~S.}\ \bibnamefont {Bolintineanu}}, \bibinfo {author} {\bibfnamefont {W.~M.}\ \bibnamefont {Brown}}, \bibinfo {author} {\bibfnamefont {P.~S.}\ \bibnamefont {Crozier}}, \bibinfo {author} {\bibfnamefont {P.~J.}\ \bibnamefont {in~’t Veld}}, \bibinfo {author} {\bibfnamefont {A.}~\bibnamefont {Kohlmeyer}}, \bibinfo {author} {\bibfnamefont {S.~G.}\ \bibnamefont {Moore}}, \bibinfo {author} {\bibfnamefont {T.~D.}\ \bibnamefont {Nguyen}}, \bibinfo {author} {\bibfnamefont {R.}~\bibnamefont {Shan}}, \bibinfo {author} {\bibfnamefont {M.~J.}\ \bibnamefont {Stevens}}, \bibinfo {author} {\bibfnamefont {J.}~\bibnamefont {Tranchida}}, \bibinfo {author} {\bibfnamefont {C.}~\bibnamefont {Trott}}, \ and\ \bibinfo {author} {\bibfnamefont {S.~J.}\ \bibnamefont {Plimpton}},\
  }\href {\doibase 10.1016/j.cpc.2021.108171} {\bibfield  {journal} {\bibinfo  {journal} {Computer Physics Communications}\ }\textbf {\bibinfo {volume} {271}},\ \bibinfo {pages} {108171} (\bibinfo {year} {2022})}\BibitemShut {NoStop}%
\bibitem [{\citenamefont {Ledbetter}\ and\ \citenamefont {Lei}(1990)}]{YBCO_bulk_modulus}%
  \BibitemOpen
  \bibfield  {author} {\bibinfo {author} {\bibfnamefont {H.}~\bibnamefont {Ledbetter}}\ and\ \bibinfo {author} {\bibfnamefont {M.}~\bibnamefont {Lei}},\ }\href {\doibase 10.1557/JMR.1990.0241} {\bibfield  {journal} {\bibinfo  {journal} {Journal of Materials Research}\ }\textbf {\bibinfo {volume} {5}},\ \bibinfo {pages} {241} (\bibinfo {year} {1990})}\BibitemShut {NoStop}%
\bibitem [{\citenamefont {Togo}\ \emph {et~al.}(2023)\citenamefont {Togo}, \citenamefont {Chaput}, \citenamefont {Tadano},\ and\ \citenamefont {Tanaka}}]{phonopyI}%
  \BibitemOpen
  \bibfield  {author} {\bibinfo {author} {\bibfnamefont {A.}~\bibnamefont {Togo}}, \bibinfo {author} {\bibfnamefont {L.}~\bibnamefont {Chaput}}, \bibinfo {author} {\bibfnamefont {T.}~\bibnamefont {Tadano}}, \ and\ \bibinfo {author} {\bibfnamefont {I.}~\bibnamefont {Tanaka}},\ }\href {\doibase 10.1088/1361-648X/acd831} {\bibfield  {journal} {\bibinfo  {journal} {Journal of Physics: Condensed Matter}\ }\textbf {\bibinfo {volume} {35}},\ \bibinfo {pages} {353001} (\bibinfo {year} {2023})}\BibitemShut {NoStop}%
\bibitem [{\citenamefont {Togo}(2023)}]{phonopyII}%
  \BibitemOpen
  \bibfield  {author} {\bibinfo {author} {\bibfnamefont {A.}~\bibnamefont {Togo}},\ }\href {\doibase 10.7566/JPSJ.92.012001} {\bibfield  {journal} {\bibinfo  {journal} {Journal of the Physical Society of Japan}\ }\textbf {\bibinfo {volume} {92}},\ \bibinfo {pages} {012001} (\bibinfo {year} {2023})}\BibitemShut {NoStop}%
\bibitem [{\citenamefont {Skelton}\ \emph {et~al.}(2016)\citenamefont {Skelton}, \citenamefont {Burton}, \citenamefont {Parker}, \citenamefont {Walsh}, \citenamefont {Kim}, \citenamefont {Soon}, \citenamefont {Buckeridge}, \citenamefont {Sokol}, \citenamefont {Catlow}, \citenamefont {Togo},\ and\ \citenamefont {Tanaka}}]{ModeMap}%
  \BibitemOpen
  \bibfield  {author} {\bibinfo {author} {\bibfnamefont {J.~M.}\ \bibnamefont {Skelton}}, \bibinfo {author} {\bibfnamefont {L.~A.}\ \bibnamefont {Burton}}, \bibinfo {author} {\bibfnamefont {S.~C.}\ \bibnamefont {Parker}}, \bibinfo {author} {\bibfnamefont {A.}~\bibnamefont {Walsh}}, \bibinfo {author} {\bibfnamefont {C.-E.}\ \bibnamefont {Kim}}, \bibinfo {author} {\bibfnamefont {A.}~\bibnamefont {Soon}}, \bibinfo {author} {\bibfnamefont {J.}~\bibnamefont {Buckeridge}}, \bibinfo {author} {\bibfnamefont {A.~A.}\ \bibnamefont {Sokol}}, \bibinfo {author} {\bibfnamefont {C.~R.~A.}\ \bibnamefont {Catlow}}, \bibinfo {author} {\bibfnamefont {A.}~\bibnamefont {Togo}}, \ and\ \bibinfo {author} {\bibfnamefont {I.}~\bibnamefont {Tanaka}},\ }\href {\doibase 10.1103/PhysRevLett.117.075502} {\bibfield  {journal} {\bibinfo  {journal} {Phys. Rev. Lett.}\ }\textbf {\bibinfo {volume} {117}},\ \bibinfo {pages} {075502} (\bibinfo {year} {2016})}\BibitemShut {NoStop}%
\bibitem [{\citenamefont {Lidiard}(1989)}]{Mott_Littleton}%
  \BibitemOpen
  \bibfield  {author} {\bibinfo {author} {\bibfnamefont {A.~B.}\ \bibnamefont {Lidiard}},\ }\href {\doibase 10.1039/F29898500341} {\bibfield  {journal} {\bibinfo  {journal} {Journal of the Chemical Society, Faraday Transactions 2: Molecular and Chemical Physics}\ }\textbf {\bibinfo {volume} {85}},\ \bibinfo {pages} {341} (\bibinfo {year} {1989})}\BibitemShut {NoStop}%
\end{thebibliography}

%

\break

\widetext
\begin{center}
\textbf{\large Supplemental Material}
\end{center}

\setcounter{equation}{0}
\setcounter{figure}{0}
\setcounter{table}{0}
\setcounter{page}{1}
\makeatletter
\renewcommand{\theequation}{S\arabic{equation}}
\renewcommand{\thefigure}{S\arabic{figure}}
\renewcommand{\bibnumfmt}[1]{[S#1]}
\renewcommand{\citenumfont}[1]{S#1}

\subsection*{Composition of data set and statistical errors from training}

Details regarding the number of configurations for each subset of configurations included in the data set, together with corresponding RMSEs for energies and forces for the training and test sets are presented in Table \ref{tab:training_set}.
The amorphous structures show the largest RMSE on both energies and forces (30 meV/atom and 350 meV/\AA), which is expected due to the very different local environments present in this set compared to the crystalline configurations.
The high RMSEs for the defect free structures are, on the other hand, due to slightly erroneous energies and forces in the data set from the constant volume relaxations, as already evidenced in the main text.

\begin{table*}[b]
\centering
\caption{Details of the data set composition in terms of type of configurations, number of atoms in each configurations, number of configurations in the set (training and test set, respectively) as well as the root mean square errors (RMSE) on energies (in meV/atom) and forces (in meV/\AA).}
\label{tab:training_set}
\begin{tabular}{ c c | c c c | c c c}
\hline
\multicolumn{2}{c}{}  & \multicolumn{3}{c}{Training} & \multicolumn{3}{c}{Test} \\
 \hline
Type & $\textrm{N}_{\textrm{atoms}}$ & $\textrm{N}_{\textrm{confs}}$ & RMSE E & RMSE F & $\textrm{N}_{\textrm{confs}}$ & RMSE E & RMSE F \\
\hline
Defect free & 416 & 468 & 14.015 & 71.482 & 26 & 12.132 & 53.187 \\
Vacancies & 415 & 404 & 2.669 & 53.561 & 20 & 3.146 & 61.416 \\
Interstitials & 417 & 127 & 1.474 & 50.923 & 6 & 2.123 & 63.410 \\
Amorphous & 52 & 62 & 30.418 & 346.052 & 3 & 25.900 & 515.334 \\
O FPs and TS & 416 & 239 & 1.062 & 68.640 & 13 & 0.713 & 54.097 \\
\hline
Total & & 1300 & 10.839 & 69.023 & 68 & 9.449 & 68.948\\    
 \hline
\end{tabular}
\end{table*}

\subsection*{Additional validation of the potential} \label{ssec:SI_validation}

We have further validated the ACE potential against quantities not directly related to the orthorhombic-to-tetragonal transition, namely bulk modulus, elastic constants, phonon dispersion relations, and formation energies of additional O FPs.
The absolute values of the equilibrium volume and lattice parameters at 0 K, bulk modulus, and elastic constants are compared in Table \ref{tab:2_properties_0K} between ACE, B+C and DFT.
DFT and B+C elastic constants are taken from Ref. \cite{GrayPotential}.
Both the ACE potential and the B+C potential predict values of bulk modulus very close to the DFT reference.
The superiority of the ACE potential is evident in the prediction of elastic constants, even though there is a slight underestimation in comparison to DFT for most components.
The largest relative discrepancy is observed for C$_{22}$, with an error of approximately 15\%.

\begin{table}
\centering
\caption{0 K equilibrium volume and lattice parameters, bulk modulus, and elastic constants of YBCO from the ACE potential, DFT, the Buckingham-Coulomb (B+C) potential of Gray et al. \cite{GrayPotential}, and experimental values.}
\label{tab:2_properties_0K}
\begin{tabular}{ c c c c c }
Property & ACE & DFT & B+C & Experiment \\
\hline
V (\AA$^3$) & 179.870 & 180.705 & 180.412 & 173.642$^a$ \\
a (\AA) & 3.859 & 3.854 & 3.877 & 3.825$^a$ \\
b (\AA) & 3.940 & 3.932 & 3.941 & 3.887$^a$ \\
c (\AA) & 11.833 & 11.924 & 11.808 & 11.678$^a$ \\
B (GPa) & 104.4 & 87.1 & 93.7 & $107 \pm 10^b$ \\
C$_{11}$ (GPa) & 197.76 & 215.97$^c$ & 166.45$^c$ & 211–234$^c$ \\
C$_{12}$ (GPa) & 96.50 & 109.24$^c$ & 73.23$^c$ & 37–132$^c$ \\
C$_{13}$ (GPa) & 50.82 & 58.03$^c$ & 42.65$^c$ & 70–100$^c$ \\
C$_{22}$ (GPa) & 199.40 & 232.40$^c$ & 180.30$^c$ & 230–268$^c$ \\
C$_{23}$ (GPa) & 52.46 & 57.16$^c$ & 52.62$^c$ & 93–100$^c$ \\
C$_{33}$ (GPa) & 142.36 & 142.56$^c$ & 163.76$^c$ & 138–186$^c$ \\
C$_{44}$ (GPa) & 49.38 & 51.19$^c$ & 48.82$^c$ & 35–61$^c$ \\
C$_{55}$ (GPa) & 41.07 & 41.19$^c$ & 28.51$^c$ & 33–50$^c$ \\
C$_{66}$ (GPa) & 84.80 & 81.73$^c$ & 69.85$^c$ & 57–97$^c$ \\
 \hline
 \multicolumn{5}{l}{\footnotesize{$^a$ Room temperature values from Ref. \cite{expTransition}.}}\\
 \multicolumn{5}{l}{\footnotesize{$^b$ Estimate from Ref. \cite{YBCO_bulk_modulus}.}}\\
 \multicolumn{5}{l}{\footnotesize{$^c$ Values reported in Ref. \cite{GrayPotential} and references therein.}}\\
\end{tabular}    
\end{table}

\begin{figure}
    \centering
    \includegraphics[width=\linewidth]{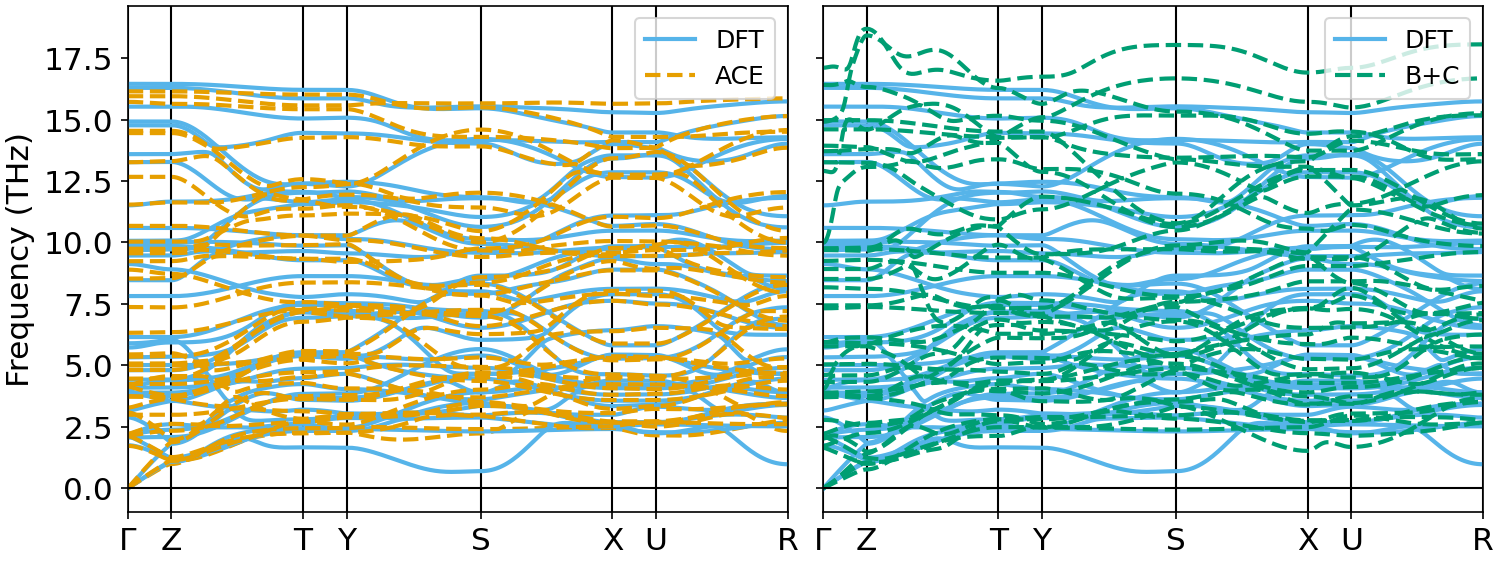}
    \caption{Phonon band structure along high symmetry paths from (left) DFT (blue solid) and ACE (orange dashed) and from (right) DFT and B+C (green dashed).}
    \label{fig:5_phonons}
\end{figure}

\begin{figure}
    \centering
    \begin{subfigure}{0.5\textwidth}
         \centering
         \includegraphics[width=\textwidth]{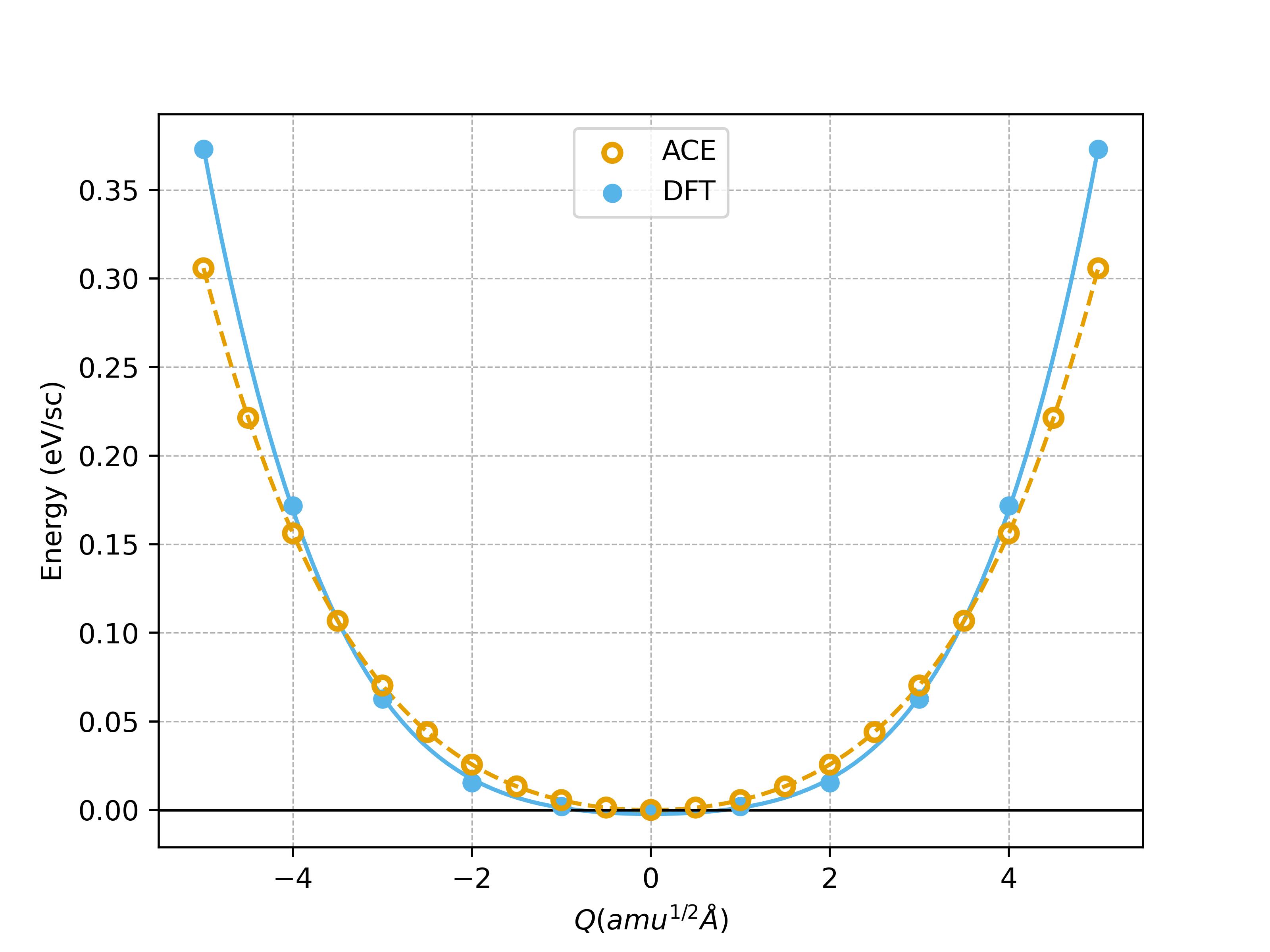}
    \end{subfigure}
    \begin{subfigure}{0.4\textwidth}
         \centering
         \includegraphics[width=\textwidth]{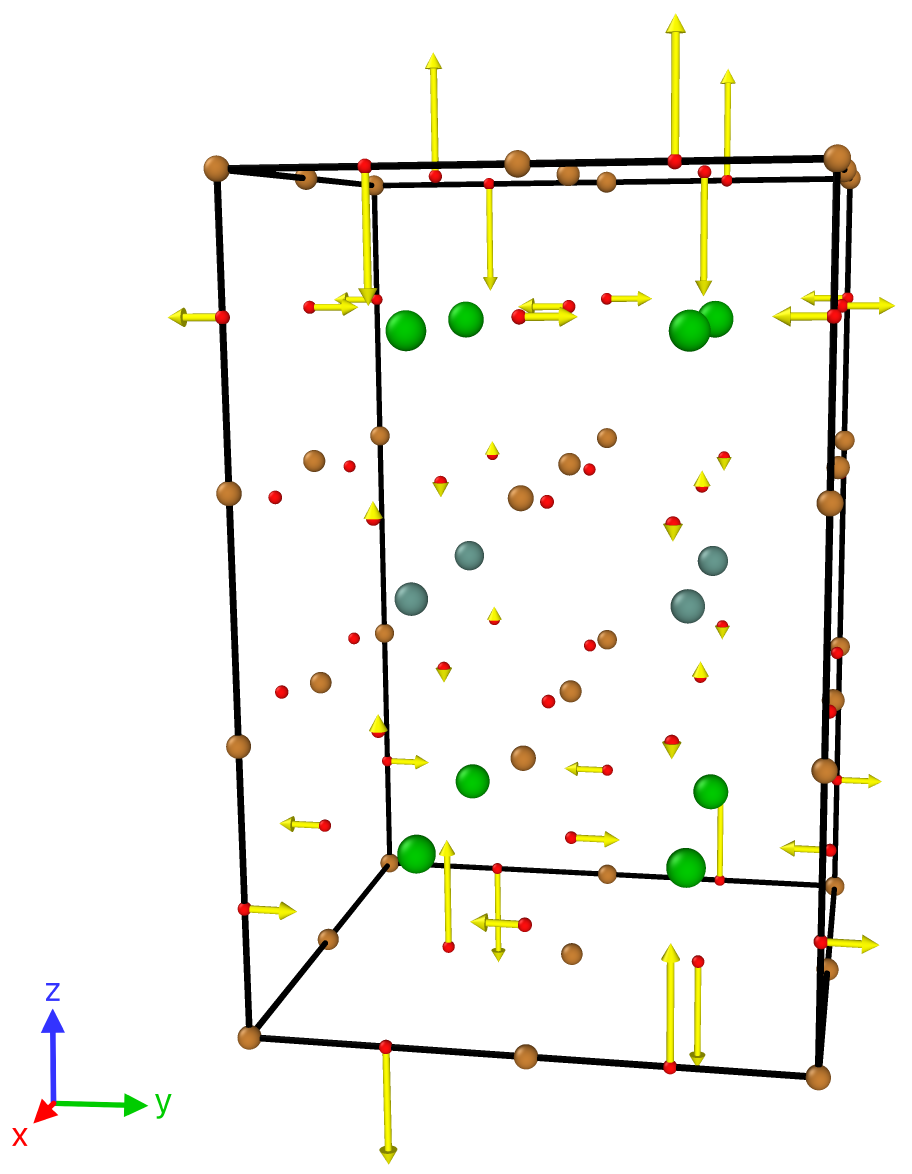}
    \end{subfigure}
    \caption{(Left) Energy landscape along the phonon mode corresponding to the lowest band at the S point from the DFT phonon dispersion calculated with DFT (blue) and ACE (orange). (Right) the atomic displacements corresponding to this phonon mode.}
    \label{fig:S1_soft_phonon}
\end{figure}

The phonon dispersion relations from the ACE and the B+C potentials are compared to DFT in Fig. \ref{fig:5_phonons}.
Phonon calculations were performed with the small displacement method as implemented in phonopy \cite{phonopyI,phonopyII} with atomic displacements of 0.01 {\AA} in a $4 \times 4 \times 2$ supercell in all cases.
Both potentials correctly predict a dynamically stable structure, but the ACE potential provides a significant improvement over the B+C potential on the optical branches, showing only minor shifts relative to DFT.
The main discrepancy is that the ACE potential does not reproduce the soft acoustic phonon mode observed in DFT around the S point.
This mismatch likely arises from the extremely small energy scales involved, which are difficult for the potential to capture with high accuracy.
Nevertheless, ACE reproduces the energies for displaced atoms along the phonon mode corresponding to the lowest band at the S point from the DFT phonon dispersion (see Fig. \ref{fig:S1_soft_phonon}) within 50 meV/supercell, equivalent to 0.1 meV/atom.
The bottom part of Fig. \ref{fig:S1_soft_phonon} shows the atomic displacements of this phonon mode (generated with ModeMap \cite{ModeMap}), which involve an anti-phase displacement of the O1 atoms in the $c$ direction, and an anti-phase displacement of the O4 atoms in the $b$ direction.
This level of accuracy suggests that the potential is able to reproduce extremely well the energetics of this phonon mode, and therefore it does not compromise the description of atomic dynamics.
It is worth noting that the DFT phonon dispersion itself required a denser k-mesh ($4 \times 4 \times 2$) than used in the training set. 
With the training k-mesh, DFT predicts the orthorhombic ground state of YBCO to be dynamically unstable.
This apparent instability probably reflects the sensitivity of the small displacement method to numerical noise: the DFT potential energy surface requires very tight convergence parameters to be smooth, whereas the ACE energy landscape is smooth by construction.

Formation energies for all O FPs and migration barriers for the most important formation mechanisms are presented in Table \ref{tab:3_Ef_and_Em_O_FPs}.
In addition to the excellent accuracy of the ACE potential in representing the O1v-O5i and O4v-O5i FPs, we observe that O2v-O5i and O3v-O5i formation energies are also well reproduced, in contrast to the poor performance of the B+C potential.
We note that the formation energies obtained here with the B+C potential differ from those reported in Ref. \cite{GrayPotential}.
This discrepancy likely arises from methodological differences: in the present work, we adopt a simple relaxation of the defect within a supercell, whereas Ref. \cite{GrayPotential} employed the Mott-Littleton approach \cite{Mott_Littleton}.

\begin{table}
\centering
\caption{Formation energies ($\textrm{E}^{\textrm{f}}$) of all O FPs considered and migration energies ($\textrm{E}^{\textrm{m}}$) for the formation of O1v-O5i and O4v-O5i  FPs from DFT, ACE potential, and B+C potential. All values are in eV. See Fig. 1 in main text for the positions of the involved atoms. Migration barriers for the formation of O2v-O5i and O3v-O5i are not reported because the physical mechanism is not likely to involve a simple migration of the atom between the initial and final sites.}
\label{tab:3_Ef_and_Em_O_FPs}
\begin{tabular}{ c | c c c | c c c} 
\hline
 & \multicolumn{3}{c}{$\textrm{E}^{\textrm{f}}$} & \multicolumn{3}{c}{$\textrm{E}^{\textrm{m}}$} \\
 \hline
O FP & DFT & ACE & B+C & DFT & ACE & B+C \\
\hline
O1v-O5i & 0.789 & 0.623 & 0.227 & 1.113 & 1.106 & 0.855 \\
O4v-O5i & 0.810 & 0.773 & -0.017 & 1.220 & 1.238 & 0.863 \\
O2v-O5i & 1.561 & 1.288 & 0.047 & - & - & - \\
O3v-O5i & 1.557 & 1.319 & 0.100 & - & - & - \\
\hline
\end{tabular}
\end{table}

\subsection*{Convergence of lattice parameters from NPT simulations}\label{ssec:SI_convergence}

\begin{figure}
    \centering
    \includegraphics[width=0.9\linewidth]{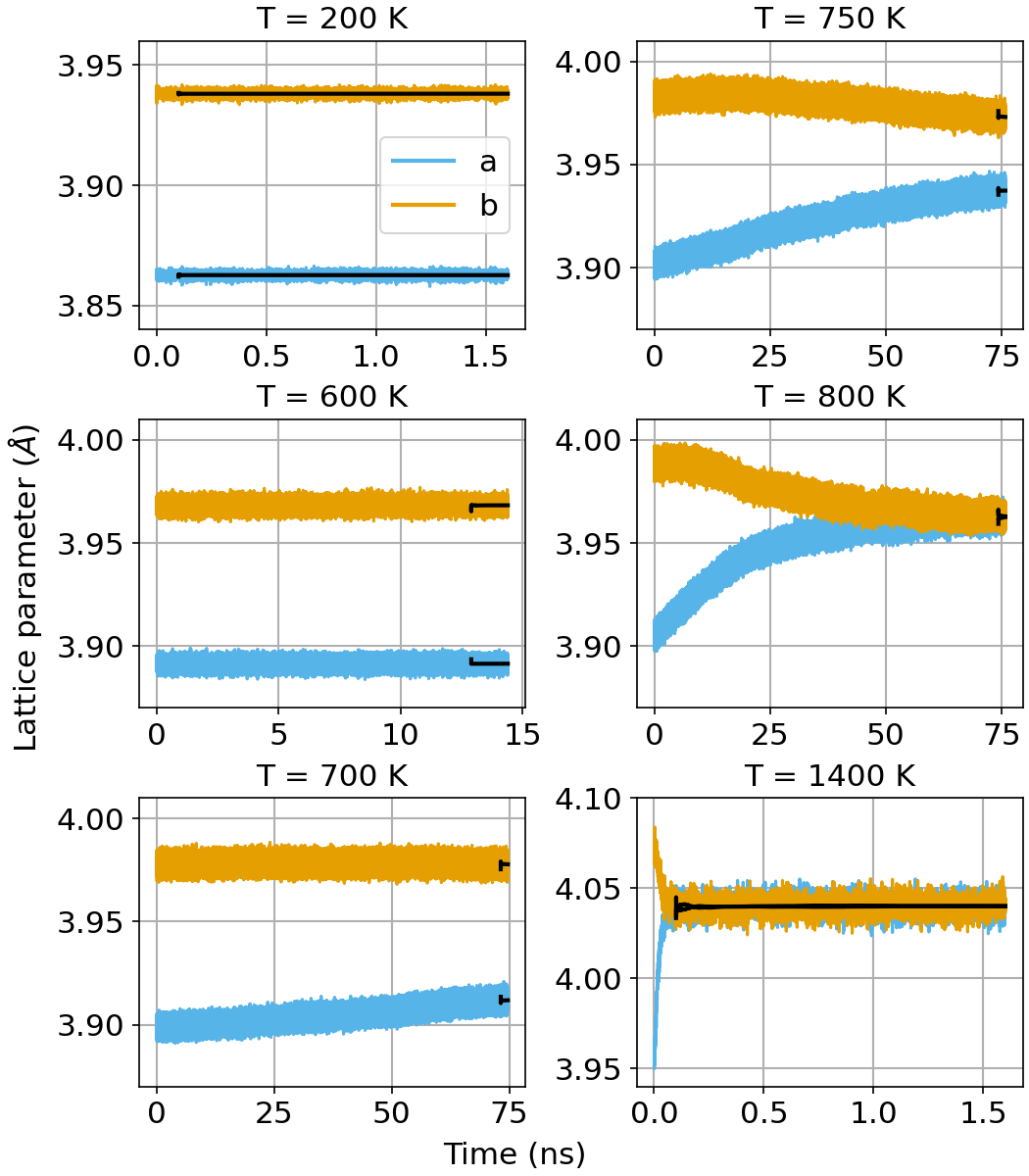}
    \caption{Convergence of the $a$ and $b$ lattice parameters as a function of time for selected temperatures. The cumulative average over the last 1.5 ns is shown in black for both parameters.}
    \label{fig:S3_a_b_convergence}
\end{figure}

The convergence of the $a$ and $b$ lattice parameters with time for selected temperatures is shown in Fig. \ref{fig:S3_a_b_convergence}.
The lowest (200 K) and highest (1400 K) temperatures are the least problematic in terms of convergence, with both lattice parameters showing stable values over the whole simulation time and cumulative averages well converged long before the end of the simulation.
At 600 K, we also observe stable values of the lattice parameters within 15 ns of simulation, even though some O5 sites start being occupied (2 sites in the whole supercell at the end of the simulation).
Between 700 and 800 K, convergence requires much longer times.
At 800 K, we needed approximately 75 ns to reach a full transition to the tetragonal structure, whereas at 700 and 750 K we can see that the $a$ and $b$ lattice are getting closer to each other but would require much longer simulation time to see if the equilibrium structure at these temperatures is the tetragonal one or a partially disordered orthorhombic structure.

\begin{figure}
    \centering
    \includegraphics[width=0.5\linewidth]{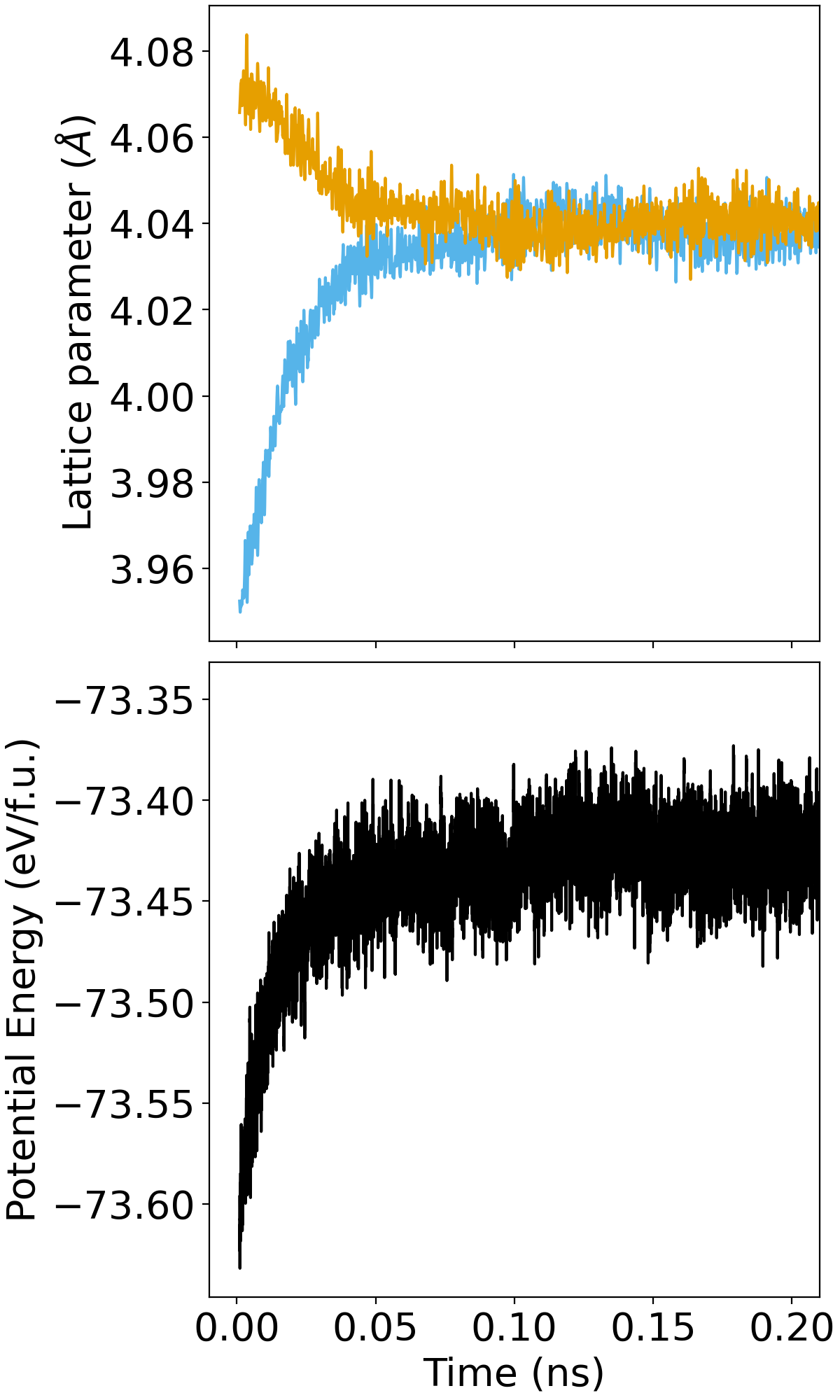}
    \caption{(Top) Convergence of the $a$ and $b$ lattice parameters as a function of time at 1400 K in the first 0.2 ns of simulation and (bottom) the associated potential energy of the system.}
    \label{fig:S4_U_vs_t_convergence}
\end{figure}

The slow dynamics of the transition could raise the question that the potential finds the tetragonal structure as more energetically favorable than the orthorhombic structure.
However, inspection of the potential energy as a function of time at 1400 K (see Fig. \ref{fig:S4_U_vs_t_convergence}) shows that the potential energy increases as the system transitions from the initial orthorhombic structure to the tetragonal one. 
This confirms that the potential correctly identifies the orthorhombic structure as the lowest-energy configuration, and that the transition to the tetragonal phase is driven by entropic effects.

\end{document}